\documentclass[hyper,a4paper]{JHEP3} 

\usepackage{epsfig}
\usepackage{latexsym}
\usepackage{graphicx}
\usepackage{amsmath}
\usepackage{amsfonts}   
\usepackage{amssymb}    

\newcommand{\newc}{\newcommand}
\newc\eg{{\it {e.g.}}}  \newc\etal{{\it {et al.}}} \newc\ie{{\it i.e.}}
\newc\etc{{\it {etc}}}  

\newcommand\lsim{\mathrel{\rlap{\lower4pt\hbox{\hskip1pt$\sim$}}
    \raise1pt\hbox{$<$}}}
\newcommand\gsim{\mathrel{\rlap{\lower4pt\hbox{\hskip1pt$\sim$}}
    \raise1pt\hbox{$>$}}}

\newc{\sigsip}{\sigma^{SI}_{p}}	\newc{\sigsin}{\sigma^{SI}_{n}}
\newc{\sigsdp}{\sigma^{SD}_{p}}	\newc{\sigsdn}{\sigma^{SD}_{n}}
\newc{\sigsi}{\sigma^{SI}}	\newc{\sigsd}{\sigma^{SD}}

\newc{\mhalf}{m_{1/2}}      \newc{\mzero}{m_0}
\newc{\tanb}{\tan\beta}
\newc{\azero}{A_0}
\newc{\at}{A_t} \newc{\abot}{A_b} \newc{\atau}{A_\tau} 

\newc{\bmu}{B\mu}           \newc{\sgn}{{\rm sgn}}
\newc{\mone}{M_1}           \newc{\mtwo}{M_2}

\newc{\charone}{\chi_1^\pm} \newc{\mcharone}{m_{\chi_1^\pm}}

\newc{\hl}{h}               \newc{\mhl}{m_{\hl}}
\newc{\hh}{H}               \newc{\mhh}{m_{\hh}}
\newc{\ha}{A}               \newc{\mha}{m_{\ha}}
\newc{\hc}{H^{\pm}}         \newc{\mhc}{m_{\hc}}

\newc{\mw}{m_{W}}      \newc{\mz}{m_{Z}}

\newc{\mgut}{M_{\rm GUT}}

\newc{\mplanck}{M_{\rm P}}      \newc{\mpl}{M_{\rm Pl}}
\newc{\msusy}{M_{\rm SUSY}}      \newc{\ms}{M_{\rm S}}

\newc{\jxf}{J({\xf})}
\newc{\jxfexact}{J_{\rm exact}({\xf})}  \newc{\jxfexp}{J_{\rm exp}({\xf})}
\newc{\VEV}[1]{\langle #1 \rangle}

\newcommand\tf{T_{\rm f}}       
\newc{\xf}{x_f}
\newc\vrel{v_{\rm rel}}
\newcommand\mchi{m_{\chi}}              

\newc\sell{{\widetilde e}_L}      \newc\msell{m_{\sell}}
\newc\selr{{\widetilde e}_R}      \newc\mselr{m_{\selr}}

\newc\snue{{\widetilde \nu}_e}      \newc\msnue{m_{\snue}}
\newc\snutau{{\widetilde \nu}_\tau}      \newc\msnutau{m_{\snutau}}

\newc\supl{{\widetilde u}_L}      \newc\msupl{m_{\supl}}
\newc\supr{{\widetilde u}_R}      \newc\msupr{m_{\supr}}
\newc\sdl{{\widetilde d}_L}      \newc\msdl{m_{\sdl}}
\newc\sdr{{\widetilde d}_R}      \newc\msdr{m_{\sdr}}

\newcommand{\stau}{{\tilde \tau}}   \newcommand\mstau{m_{\stau}}
\newcommand{\stauone}{{\tilde \tau}_1}   \newcommand\mstauone{m_{\stauone}}

\newcommand\gluino{\tilde g}
\newcommand\mgluino{m_{\gluino}}

\newc\hpm{H^\pm} \newc\hp{H^+} \newc\hm{H^-} 
\newc\sfermion{\tilde f}  \newc\msfermion{m_{\sfermion}}  

\newc\alphas{\alpha_s}

\newc\alphaem{\alpha_{em}}

\newcommand\treh{T_{\rm R}}     \newcommand\trehmax{T_{\rm R}^{\rm max}}

\newc{\sthw}{\sin\theta_W}       \newc{\ssqthw}{\sin^2\theta_W}         
\newc{\cthw}{\cos\theta_W}       \newc{\csqthw}{\cos^2\theta_W}
\newc{\tthw}{\tan\theta_W}       \newc{\tsqthw}{\tan^2\theta_W}       

\newc{\bino}{\widetilde B}              \newc{\wino}{\widetilde W_3}
\newc{\higgsinob}{{\widetilde H}^0_b}   \newc{\higgsinot}{{\widetilde H}^0_t}

\newc{\abund}{\Omega h^2}               \newc{\abundobs}{\Omega_{\rm obs} h^2}

\newc{\abundchi}{\Omega_\chi h^2}
\newc{\abundcdm}{\Omega_{{\rm CDM}} h^2}

\newc{\omegam}{\Omega_{{\rm M}}}       \newc{\abundm}{\Omega_{{\rm M}} h^2}
\newc{\omegab}{\Omega_{{\rm b}}}	\newc{\abundb}{\Omega_{{\rm b}} h^2}
\newc{\omegatot}{\Omega_{{\rm TOT}}}

\newc{\nlsp}{n_{{\rm LSP}}} \newc{\mlsp}{m_{{\rm LSP}}} \newc{\mlspmax}{m_{\rm LSP}^{\rm max}}     
\newc{\ylsp}{Y_{{\rm LSP}}} 
\newc{\abundlsp}{\Omega_{\rm LSP}h^2}
\newc{\abundlsptp}{\Omega_{\rm LSP}^{\rm TP}h^2} 
\newc{\abundlspntp}{\Omega_{\rm LSP}^{\rm NTP}h^2}

\newc{\omeganlsp}{\Omega_{{\rm NLSP}}}   
\newc{\ynlsp}{Y_{{\rm NLSP}}}            \newc{\taunlsp}{\tau_{{\rm NLSP}}}
\newc{\nnlsp}{n_{{\rm NLSP}}}            \newc{\mnlsp}{m_{{\rm NLSP}}}
\newc{\abundnlsp}{\Omega_{\rm NLSP}h^2}
\newc{\abundnlsptp}{\Omega_{\rm NLSP}^{\rm TP}h^2} 
\newc{\abundnlspntp}{\Omega_{\rm NLSP}^{\rm NTP}h^2}

\newc{\nx}{n_{X}}                        \newc{\yx}{Y_{X}}
\newc{\mx}{m_{X}}                        \newc{\taux}{\tau_{X}}

\newc{\rhocrit}{\rho_{crit}}
\newc{\rhochi}{\rho_{\chi}}

\newcommand\fa{f_{a}}

\newcommand\neut{\tilde \chi}

\newc{\cachigamma}{C_{a\neut\gamma}}
\newc{\caww}{C_{aWW}}                   
\newc{\cayy}{C_{aYY}}

\newc{\nl}{\cos \theta_{\tilde t}}
\newc{\nr}{\sin \theta_{\tilde t}}
\newcommand\tev{\,{\rm TeV}}
\newcommand\gev{\,{\rm GeV}}

\newcommand\kev{\,{\rm keV}}

\newcommand\pb{\,\mbox{pb}}

\newc\gbar{{\overline{g}}}

\newc{\ra}{\rightarrow}

\newc{\beq}{\begin{equation}}
\newc{\eeq}{\end{equation}}
\newc{\bea}{\begin{eqnarray}}
\newc{\eea}{\end{eqnarray}}
\newcommand{\beqa}[1]{\begin{eqnarray}#1\end{eqnarray}}

\newc{\nspin}{n_{\rm spin}}
\newc{\nflavor}{n_{\rm F}}
\newc{\ngamma}{n_\gamma}
\newc{\ychi}{Y_{\chi}}                  \newc{\yeqchi}{Y^{\rm EQ}_{\chi}}

\newcommand\axino{{\tilde{a}}}        
\newcommand\maxino{{m_{\axino}}}

\newcommand\abunda{\Omega_{\axino}h^2}
\newcommand\abundantp{\Omega^{\rm NTP}_{\axino}h^2}     

\newcommand\abundatp{\Omega^{\rm TP}_{\axino}h^2}       

\newc{\naxino}{n_{\axino}}
\newc{\yaxino}{Y_{\axino}}
\newc{\yaxinoeq}{Y^{\rm EQ}_{\axino}}
\newc{\yaxinotp}{Y^{\rm TP}_{\axino}}
\newc{\yaxinontp}{Y^{\rm NTP}_{\axino}}

\newcommand\gravitino{{\widetilde{G}}}    
\newcommand\mgravitino{{m_{\gravitino}}}

\newcommand\abundg{\Omega_{\gravitino}h^2}

\newcommand\abundgtp{\Omega^{\rm TP}_{\gravitino}h^2}       

\newc{\ngravitino}{n_{\gravitino}}
\newc{\ygravitino}{Y_{\gravitino}}
\newc{\yeqgravitino}{Y^{\rm EQ}_{\gravitino}}
\newc{\ygravitinotp}{Y^{\rm TP}_{\gravitino}}
\newc{\ygravitinontp}{Y^{\rm NTP}_{\gravitino}}

\newc{\yascat}{Y^{\rm scat}_{i,j}}      \newc{\yadec}{Y^{\rm dec}_{i}}
\newc{\gstar}{g_\ast}           \newc{\gsstar}{g_{s\ast}}

       \def\pslash{\not{\hbox{\kern-2.3pt $p$}}}
       \def\kslash{\not{\hbox{\kern-2.3pt $k$}}}
       \def\qslash{\not{\hbox{\kern-2.3pt $q$}}}
       \def\ddslash{\not{\hbox{\kern-2.3pt $d$}}}
       \def\prtslash{\not{\hbox{\kern-2.3pt $\partial$}}}

\title{Determining Reheating Temperature at Colliders\\ with Axino
  or Gravitino Dark Matter}
\author{Ki-Young Choi\\
        Department of Physics and Astronomy, University of Sheffield,\\
        Sheffield S3 7RH, England, and\\
        Departamento de F\'{\i}sica Te\'{o}rica C-XI
        and Instituto de F\'{\i}sica Te\'{o}rica UAM/CSIC,\\
        Universidad Aut\'{o}noma de Madrid, Cantoblanco,
        28049 Madrid, Spain\\
        E-mail: \email{K.Choi@sheffield.ac.uk}}
\author{Leszek Roszkowski\\
        Department of Physics and Astronomy, University of Sheffield,\\
        Sheffield S3 7RH, England, and\\
	Theory Division, CERN, CH-1211 Geneva 23, Switzerland\\
        E-mail: \email{L.Roszkowski@sheffield.ac.uk}}
\author{Roberto Ruiz de Austri\\
        Departamento de F\'{\i}sica Te\'{o}rica C-XI
        and Instituto de F\'{\i}sica Te\'{o}rica UAM/CSIC,\\
        Universidad Aut\'{o}noma de Madrid, Cantoblanco,
        28049 Madrid, Spain\\
        E-mail: \email{rruiz@delta.ft.uam.es}}

\abstract{\small After a period of inflationary expansion, the
  Universe reheated and reached full thermal equilibrium at the
  reheating temperature $\treh$. In this work we point out that, in
  the context of effective low-energy supersymmetric models, LHC
  measurements may allow one to determine $\treh$ as a function of the
  mass of the dark matter particle assumed to be either an axino or a
  gravitino. An upper bound on their mass and on $\treh$ may also be
  derived.}

\begin{document}


\section{Introduction}\label{sect:intro}

Dark matter (DM) remains an unknown component of the Universe. While
it has so far escaped detection, its existence has been convincingly
inferred from gravitational effects that it imparts on visible matter
through rotational curves of spiral galaxies, gravitational lensing,
etc,~\cite{susy-dm-reviews}. The effects of dark matter also can be
seen on large structure formation and on anisotropy of the cosmic
microwave background (CMB).  The CMB in particular provides a powerful
tool for determining the global abundance of cold DM (CDM).  Recently,
the Wilkinson Microwave Anisotropy Probe (WMAP) has performed a
high-accuracy measurement of several cosmological parameters. In
particular, the relic density of CDM has been determined to lie in the
range~\cite{Spergel:2006hy}
\beq
\abundcdm =0.104 \pm 0.009. 
\label{Oh2WMAP}
\eeq
Since DM has to be electrically and (preferably) color-charge neutral,
from the particle physics point of view, a natural candidate for DM is
some weakly interacting massive particle (WIMP). Within standard
cosmology, the WIMP is produced via a usual freeze-out
mechanism from an expanding plasma.  

Among specific, well-motivated particle candidates for the WIMP, the
by far most popular choice is a stable lightest neutralino $\chi$ of
effective low-energy supersymmetry (SUSY) models. Most efforts have
gone to exploring the lightest neutralino $\chi$ of the Minimal
Supersymmetric Standard Model (MSSM) as the lightest supersymmetric
particle (LSP) that, in the presence of R-parity, is stable and makes
up all, or most of, the CDM in the Universe.  In addition to an
impressive experimental effort of direct and indirect searches for
cosmic WIMPs, the Large Hadron Collider (LHC) will soon start
exploring the TeV energy scale and is expected to find several
superpartners and to determine their properties. In particular, some
authors have explored the feasibility of determining the neutralino's
relic abundance $\abundchi$ from LHC
measurements~\cite{oh2atlhc-cmssm,oh2atlhc-mssm}.  Their conclusion
was that, under favorable circumstances, this should be possible with
rather good accuracy, of order 10\% or better, although this may be
challenging~\cite{bk05}. An analogous study has also been done in the
context of the Linear Collider, where accuracy of a similar
determination would be much better~\cite{oh2atilc}.

If a WIMP signal is detected in one or more DM detection experiments,
and also at the LHC a large missing-mass and missing-energy signature,
characteristic of the stable neutralino, is measured and implies a
similar mass, and if perhaps additionally eventually its relic
abundance is determined from LHC data, even if with limited precision,
and agrees with the ``WMAP range''~(\ref{Oh2WMAP}), then the DM
problem will most likely be declared solved, and for a good reason.

However, such an optimistic outcome is by no means guaranteed. One
realistic possibility is that the neutralino, even if it is found as
an apparently stable state in LHC detectors, may not be the true LSP
and therefore DM in the Universe. Instead, it could decay in the early
Universe into an even lighter, and (possibly much) more weakly
interacting, state, the real LSP, outside the MSSM (or some other
low-energy SUSY model) spectrum. In this case, current cosmic WIMP
searches will prove futile, even after improving the upper limit on
the spin-independent interaction cross section on a free proton
$\sigsip$ from the current sensitivity of
$\sim10^{-7}\pb$~\cite{silimit-07} down to $\sim10^{-10}\pb$, which is as far down as
experiment can probably go given background from natural
radioactivity.

Moreover, the neutralino relic abundance, as determined at the LHC,
may come out convincingly outside the range~(\ref{Oh2WMAP}).  In fact,
a value of $\abundchi$ above about 0.1 can be easily explained in
terms of a lighter LSP into which the neutralino, after its
freeze-out, decayed in the early Universe~\cite{ckr}. This is because
the relic abundance of the true LSP is in this case related to
$\abundchi$ by the ratio of the LSP mass $\mlsp$ to $\mchi$, which is
less than one. If $\abundchi$ comes out below~(\ref{Oh2WMAP}), several
solutions have been suggested which invoke non-standard cosmology,
e.g. quintessence-driven kination~\cite{lowoh2-quint}, while
preserving the neutralino as the DM in the Universe.  However, if at
the same time DM searches bring null results, this will provide a
strong indication against the neutralino nature of DM.  In contrast to
the above attempts, we will consider a whole range of possible values
of $\abundchi$ at the LHC, both below and above 0.1. We will work
instead within the framework of standard cosmology but  will
not assume the neutralino to be the DM in the Universe.

In this context we point an intriguing possibility of determining at
the LHC the reheating temperature of the Universe. The framework we
consider will therefore give us an opportunity to probe some crucial
features of the early period of the Universe's history.  The reheating
temperature $\treh$ is normally thought of as the temperature at
which, after a period of rapid inflationary expansion, the Universe
reheated (or, more properly, defrosted), and the expanding plasma
reached full thermal equilibrium. Determining $\treh$ cannot be done
with the neutralino as DM since it freezes out at $\tf\simeq
\mchi/24$, which is normally thought to be much below
$\treh$.\footnote{The possibility of $\treh\lsim\tf$ has been explored
in~\cite{gkr00} and more recently in~\cite{ggetal06}. In this case
one could think of determining experimentally
$\treh$ even with the neutralino as DM.}  Here instead we assume a
different candidate for the LSP (assuming R-parity) and cold DM, whose
relic density depends, at least in part, on $\treh$. This is the case
for either an axino or a gravitino.  The spin-$1/2$ axino (the
fermionic superpartner of an axion) and the spin-$3/2$ gravitino (the
fermionic superpartner of a graviton) are both well-motivated. The
former arises in SUSY extensions of models incorporating the
Peccei-Quinn solution to the strong CP problem. The latter is an
inherent ingredient of the particle spectrum of supergravity models.
Both, like the axion, form a subclass of extremely weakly interacting
massive particles (E-WIMPs)~\cite{Choi:2005vq}.  The characteristic
strength of their interactions with ordinary matter is strongly
suppressed by a large mass scale, the Peccei-Quinn scale $\fa\sim
10^{11}\gev$ in the case of axinos and the (reduced) Planck scale
$\mplanck\simeq 2.4\times 10^{18} \gev$ for gravitinos. The mass of
the axino $\maxino$ is strongly model-dependent and can take values
ranging from keV up to TeV~\cite{ckn}. The mass of the gravitino
$\mgravitino$ in gravity-mediated SUSY breaking schemes is given by
$\mgravitino\sim{\ms^2/\mplanck}$, where $\ms\sim10^{11}\gev$ is the
scale of local SUSY-breaking in the hidden sector, and is expected in
the GeV to TeV regime. In other schemes of SUSY breaking $\mgravitino$
can be (much) smaller. In this work we want to remain as
model-independent as possible and will treat $\maxino$ and
$\mgravitino$ as free parameters.  Both axinos and gravitinos can be
produced in decays of the neutralino (or another ordinary
superpartner, e.g. the stau) after freeze-out, as mentioned above, or
in thermal scatterings and decay processes of ordinary particles and
sparticles in hot plasma at high enough reheating temperatures $\treh$
- hence their relic density dependence on $\treh$. The possibility of
axinos in a KSVZ axion framework~\cite{ksvz} as cold DM was pointed
out in~\cite{ckr,ckkr} and next studied in several
papers~\cite{crs02,crrs04,bs04,rs06}, while axinos as warm DM was
considered in~\cite{rtw}. The gravitino as a cosmological relic was
extensively studied in the literature, starting
from~\cite{gravitino-early,ekn84}, more recent papers on gravitino CDM
include~\cite{mmy93,bbp98,bbb00,fengetal,rrc04,ccjrr}.

For definiteness, in this work we will first assume the lightest
neutralino to be the lightest ordinary superpartner and the
next-to-lightest particle (NLSP), although below we will also consider
the case of the stau. Our main result is that, assuming the axino or
gravitino as the true LSP and CDM, and that the (apparently stable)
neutralino is discovered at the LHC and its ``relic abundance''
$\abundchi$ is determined from LHC data, then one should be able to
determine the reheating temperature in the Universe as a function of
the LSP mass, or at least place an {\em upper bound} on it, as we show below.
In the regime where thermal
production dominates, in the axino case we find $\treh\propto
\fa^2/\maxino$ while in the gravitino case $\treh\propto
\mgravitino$. Alternatively, in the non thermal production dominated
regime, we find an {\em upper bound} on the allowed mass range of a
CDM particle.
Furthermore, $\abundchi$ at the LHC can be expected to come out either
below or above (or for that matter even accidentally agree with) the
``WMAP range''~(\ref{Oh2WMAP}).

The same holds true also in the even more striking case of the
lighter stau taking the role of the NLSP instead. In this case the very discovery
of an (apparently stable) charged massive particle at the LHC will
immediately imply that DM is made up of some state outside the usual
spectrum of low-energy superpartners. In this case, dedicated
studies of the differential photon spectrum may allow one to
distinguish between the axino and the gravitino LSP~\cite{bchrr05}.

We stress that, a detection of a (seemingly) stable neutralino at
the LHC will not be sufficient to prove that the neutralino is the LSP
and the CDM.\footnote{Note that, for the neutralino, or any other
superpartner, to appear stable in an LHC detector, it is sufficient
that its lifetime is longer than a microsecond or so. Of course, if it
is unstable, then it will be immediately clear that the neutralino is
not cosmologically relevant.}  Even establishing an apparent agreement
of $\abundchi$, as to be determined at the LHC, with the range~(\ref{Oh2WMAP})
will not necessarily imply the neutralino nature of DM, although
admittedly it will be very persuasive.  For this, a
signal in DM searches must also be detected and the resulting WIMP
mass must be consistent with LHC measurements of the neutralino mass.

The paper is organized as follows. In section~\ref{production},
we review the thermal and non-thermal production of axinos and
gravitinos and next explain our strategy for determining $\treh$ at the
LHC.  In sections~\ref{axino} and~\ref{gravitino}, we discuss the
reheating temperature determination with axino LSP and gravitino LSP
respectively, assuming the neutralino to be the NLSP. In
section~\ref{staunlsp} we consider instead the lighter stau as the
NLSP. We make final remarks and summarize our findings in section~\ref{summary}.

\section{Cosmological production of relic axino or gravitino E-WIMPs}
\label{production}

First we briefly review main mechanisms of producing relic axino or
gravitino E-WIMPs, assumed to be the LSP, in the early Universe.
Since both are neutral Majorana particles, they are produced in an
analogous way. Since the interactions of axino and gravitino are strongly
suppressed with respect to the Standard Model interaction strengths by
$\fa$ and $\mplanck$, respectively, 
they can be in thermal equilibrium only at very high temperatures of
order $10^{11}\gev$.  Axinos and/or gravitinos produced at lower
temperatures are out-of-equilibrium, thus their abundance depends on
the reheating temperature.

Neglecting the initial equilibrium population (which we assume to have
been diluted through inflation), we consider here two generic ways to
re-populate the Universe with axinos or gravitinos.  One proceeds via
scatterings and decay processes of ordinary particles and sparticles
in thermal bath. Its efficiency is proportional to their density in
the plasma which is a function of $\treh$. Following~\cite{ckkr} we
call it {\em thermal production} (TP). The other, dubbed {\em
non-thermal production} (NTP) refers to (out-of-equilibrium) decays of
the NLSPs, after their freeze-out, to E-WIMPs.  In addition, there could be
other possible mechanisms contributing to E-WIMP population,
e.g. through inflaton decay, but they are much more model dependent
and will not be considered here.

\vspace*{\baselineskip}
\noindent {\bf Thermal production at high {\boldmath
$\treh$}.}\hspace*{0.3cm} In considering thermal production of E-WIMPs
we compute their yield $Y^{\rm TP}_{\rm LSP} \equiv n_{\rm LSP}/s$,
where $n_{\rm LSP}$ is their number density and $s$ is the entropy
density. To this end, we integrate the Boltzmann equation up to the
reheating temperature $\treh$.  Their thermal abundance is then given
by
\beqa{
\abundlsptp = \mlsp Y^{\rm TP}_{\rm LSP} \frac{s h^2}{\rhocrit},
\label{eq:defabund_tp}
}
where $h$ is the dimensionless Hubble parameter and $\rhocrit$ is the
critical density. In other words, $\mlsp Y^{\rm TP}_{\rm
LSP}=3.7\times10^{-10}\left(\abundlsptp/0.1 \right)$, which is
actually true in general.

In the case of axinos we follow the procedure described in detail
in~\cite{ckr,ckkr,crs02,crrs04}. The main production channels are the
scatterings of strongly interacting (s)particles described by the dimension-five
axino-gluino-gluon term 
\beqa{
{\cal L}_\axino \owns i\frac{\alphas}{16\pi \fa}
\bar{\axino}\gamma_5\left[\gamma^\mu,\gamma^\nu\right] \gluino^a F^a_{\mu\nu}.
\label{eq:axinolang}
}
The most important contributions come from 2-body processes into final
states, $i+ j\rightarrow
\axino+\cdots$~\cite{ckr,ckkr}.  Axinos also can be
produced through the decays of heavier superpartners in thermal
plasma. In this work we include gluino, squark, slepton and neutralino
decays.  At low reheating temperatures, comparable to the mass of the
decaying particle, axino production from squark decay in thermal bath
gives dominant contribution~\cite{ckkr}.

\begin{figure}[!tb]
  \begin{center}
  \begin{tabular}{c c}
    \includegraphics[width=0.45\textwidth]{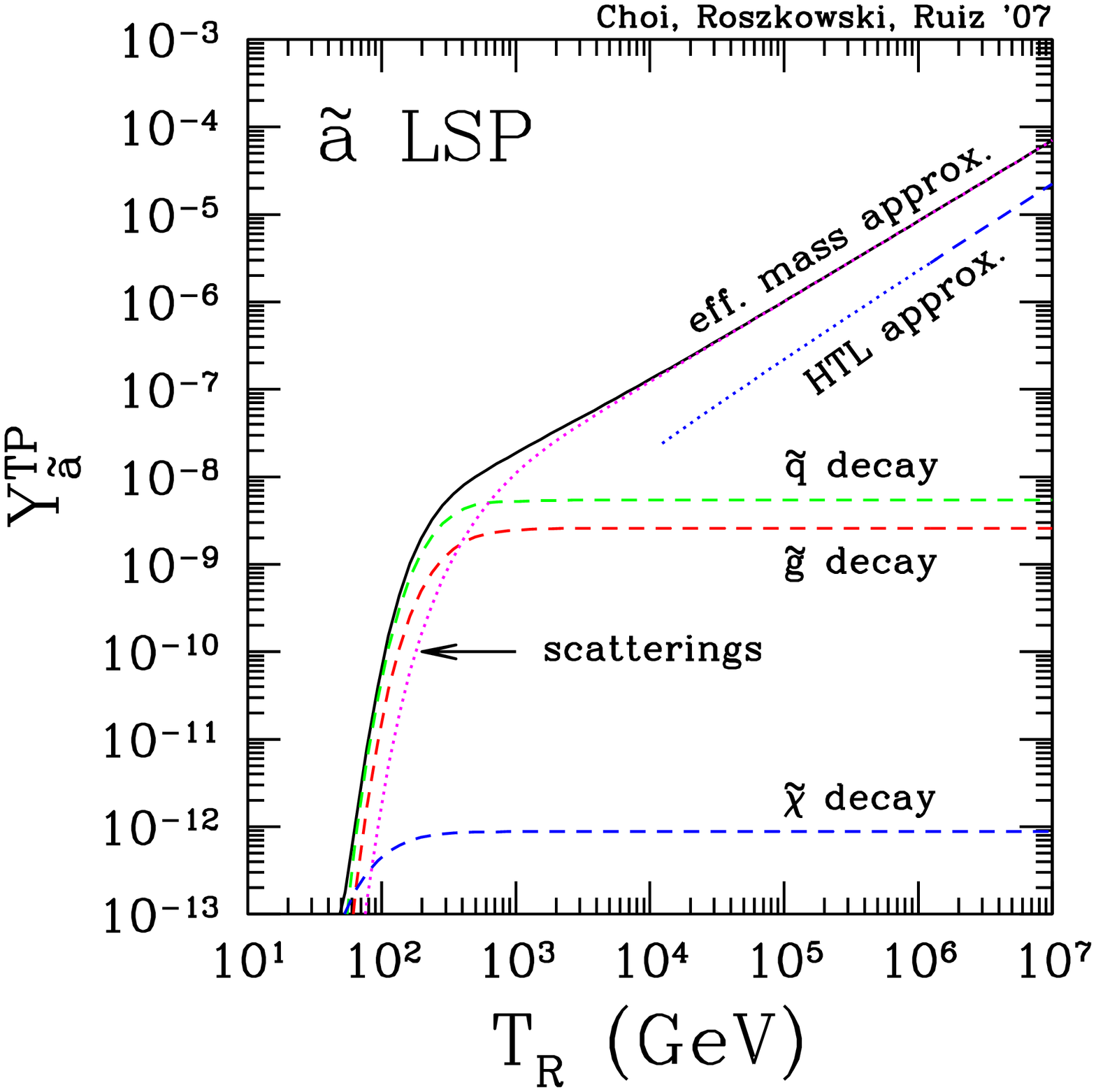}
&
    \includegraphics[width=0.45\textwidth]{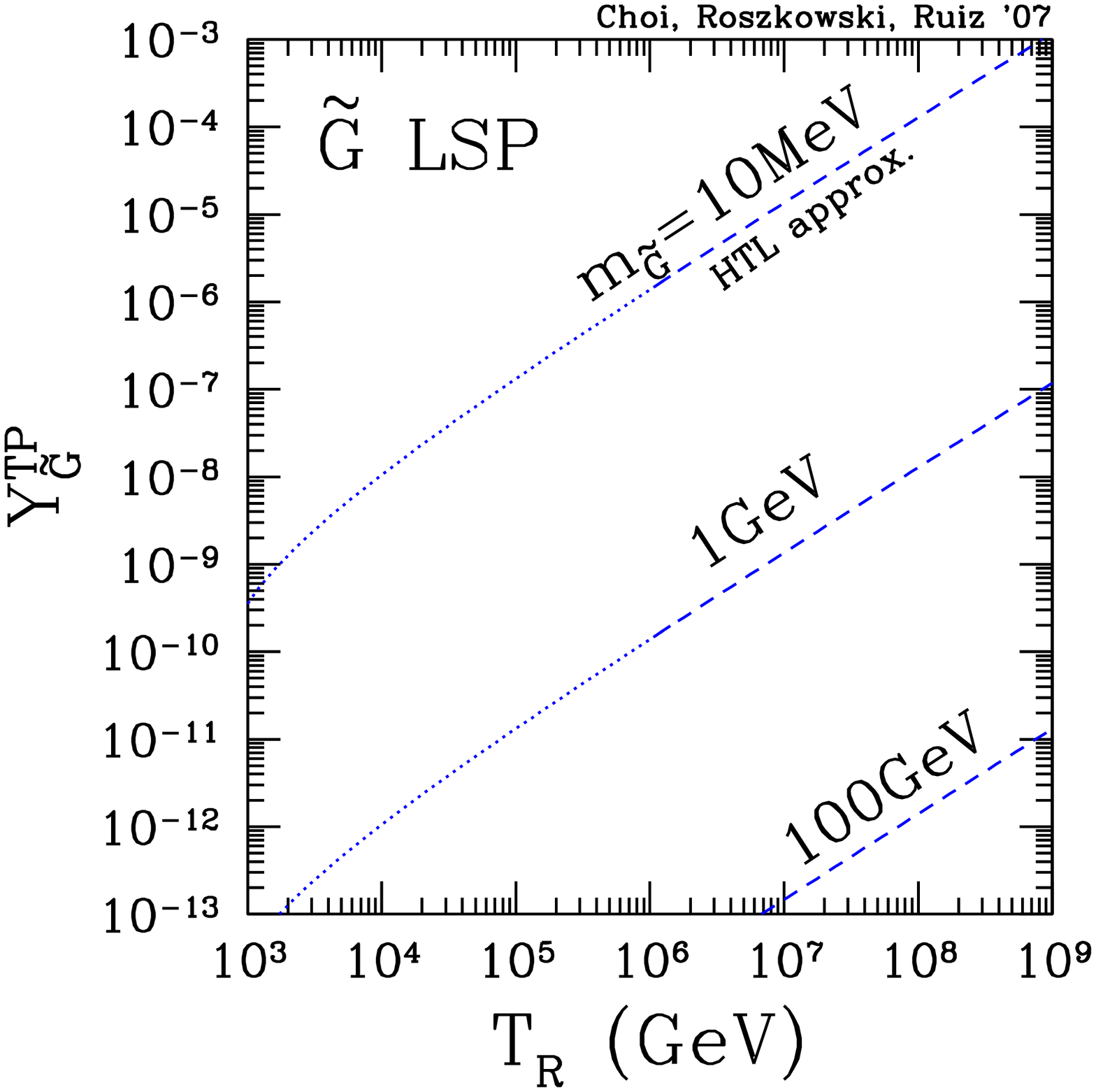}
    \end{tabular}
  \end{center}
\caption{The thermal yield of axinos $\yaxinotp$ (left
  panel) and gravitinos $\ygravitinotp$ (right panel) as a function of
  $\treh$. In both panels the neutralino mass is $\mchi=300 \gev$, the
  stau mass is $\mstauone=474 \gev$, while gluino, squark masses are
  set at $1\tev$. In the left panel we take $\fa=10^{11} \gev$; the
  yield scales as $1/\fa^2$. The black solid curve has been obtained
  within the effective (plasmon) mass approximation, as implemented
  in~\cite{ckkr}, which we use in this work. For comparison, the blue
  curve has been obtained in~\cite{bs04} by applying a hard thermal
  loop (HTL) technique. We also show the relative contributions to
  $\yaxinotp$ from squark, gluino and neutralino decays. In the right
  panel we compute the gravitino yield following ref.~\cite{bbb00}
  where the HTL technique was applied. In both panels the results
  obtained within the HTL formalism are more correct at
  $\treh\gsim10^6\gev$ (dashed) but less applicable at lower $\treh$
  (dotted).  }
\label{fig:Y_TP}
\end{figure}

In the gravitino case, the analogous dominant goldstino-gluino-gluon
dimension-five term is given by~\cite{mmy93,bbb00}
\beqa{
{\cal L}_\gravitino \owns \frac{\mgluino}{\sqrt{2}6\mplanck \mgravitino}
\bar{\psi}_G\left[\gamma^\mu,\gamma^\nu\right] \gluino^a F^a_{\mu\nu},
\label{eq:gravitinolang}
}
where the goldstino ${\psi}_G$ is the spin-$1/2$ component of the
gravitino. Note the dependence on $\mgluino/\mgravitino$, while no
analogous term appears in the axino case.  In computing the gravitino
yield from thermal processes we follow ref.~\cite{bbb00}.  On the
other hand, we do not include gravitino production from sparticle
decays in the plasma~\cite{rs06} which change the yield by a factor of
about two.

Fig.~\ref{fig:Y_TP} shows the yield of axinos $\yaxinotp$ (left panel)
and of gravitinos $\ygravitinotp$ (right panel) from thermal
production as a function of $\treh$.  Input parameters are given in
the caption. As expected, in both cases the yield grows with
increasing $\treh$. In the axino case, a sharp drop-off below
$\treh\sim1\tev$ is due to Boltzmann suppression factor
$\exp{(-m/T)}$, with $m$ denoting here squark and gluino mass; at lower $\treh$
superpartner decay processes become dominant but are less
efficient~\cite{ckkr}.

In both axino and gravitino production from scatterings in thermal
plasma, some classes of diagrams suffer from infra-red divergence. An
early remedy in terms of a plasmon mass as a infra-red regulator in
the gravitino case in~\cite{ekn84} was more recently improved using
hard thermal loop (HTL) technique and a finite result was
obtained~\cite{bbp98,bbb00}. The HTL technique is fully applicable
only in cases corresponding to rather high $\treh\gsim10^6\gev$ for
which, in the right panel of fig.~\ref{fig:Y_TP}, the gravitino yield
is marked with a dashed line for a few representative choices of
$\mgravitino$. At lower $\treh$ we use the same formalism but the
result is less reliable (dotted parts).  In the axino case an
analogous calculation was performed in~\cite{bs04} but is of more
limited use since for the axino to be CDM one requires
$\treh\lsim10^6\gev$~\cite{ckkr}.  In this work in computing the axino
yield we therefore follow~\cite{ckkr} and use the plasmon mass as a
regulator.  The result is shown as solid line in the left panel of
fig.~\ref{fig:Y_TP} and labelled ``effective mass approximation''. For
comparison, the blue line shows the result obtained in~\cite{bs04}
using the HTL technique which is more correct than our treatment in
the regime marked by a dashed line but less reliable in the one marked
by a dotted one ($\treh\lsim10^6\gev$). In any case, the difference is
of order a few which is not important for our purpose.

Since the axino yield from thermal production $\yaxinotp$ is basically
independent of the axino mass, until it becomes comparable to the masses
of MSSM sparticles, its relic abundance $\abundatp$ is proportional to
$\maxino$ and can then be expressed as~\cite{ckkr}
\beqa{
\abundatp =0.1\left(\frac{\maxino}{100\gev} \right) 
\left(\frac{\yaxinotp}{3.7\times 10^{-12}} \right).
\label{eq:TP_axino}
}
At high enough $\treh$ thermal production is dominated by scattering
processes and an application of the HTL leads to the following
formula~\cite{bs04}
\beqa{
\abundatp\simeq 5.5\, g_s^6 \ln \left(\frac{1.108}{g_s} \right)
\left(\frac{\maxino}{0.1\gev} \right)
\left(\frac{10^{11}\gev}{\fa}\right)^2 \left(\frac{\treh}{10^4 \gev} \right),
\label{eq:TP_axino_bs}
}
where $g_s$ is temperature-dependent strong coupling constant, which
in the above expression is evaluated at $\treh$. Note that,
$\yaxinotp\propto\treh/\fa^2$, as expected.

In the gravitino case, due to the above-mentioned
gravitino mass dependence in the denominator of the dimension-five
terms of the gravitino Lagrangian, $\ygravitinotp\propto
1/\mgravitino^2$, and the relic abundance calculated using a HTL
technique can be expressed as~\cite{bbb00}
\begin{equation}
\abundgtp\simeq 0.27 \left(\frac{\treh}{10^{10}\gev}\right)
\left(\frac{100\gev}{\mgravitino}\right) 
\left(\frac{\mgluino(\mu)}{1\tev}\right)^2,
\label{eq:abundgbbb}
\end{equation}
where $\mgluino(\mu)$ stands for the gluino
mass evaluated at a scale $\mu\simeq 1\tev$. 

\vspace*{\baselineskip}
\noindent {\bf Non-thermal production from NLSP
decays.}\hspace*{0.3cm} 
As the Universe cools down, all heavier
superparticles first cascade decay into the lightest superpartners of
the low-energy SUSY spectrum, in our case the NLSPs. The NLSPs then
freeze out from thermal equilibrium and only later decay to LSP axinos
or gravitinos.  Since all the NLSPs decay into axino or gravitino
LSPs, their number density is the same as that of decaying NLSP,
$\ylsp=\ynlsp$, and the LSP relic abundance from non-thermal
production is given by
\beq
\abundlspntp = \frac{\mlsp}{\mnlsp} \abundnlsp.
\label{eq:ntp}
\eeq

Once supersymmetric particles are found at the
LHC and relevant parameters of the NLSP are determined, then its
freeze-out relic abundance can be calculated by solving the Boltzmann
equation (under the assumption that it is actually stable until after
its freeze-out).\footnote{Obviously, in our case $\abundnlsp$ is not
the true cosmological relic abundance of the NLSPs. Instead it is
simply given by $\abundnlsp=\mnlsp\ynlsp \frac{s h^2}{\rhocrit}$ where
$\ynlsp$ is the NLSP yield at its freeze-out. Here, for convenience we
will keep calling $\abundnlsp$ a ``relic abundance''.} Since the NLSPs do
not constitute CDM, we treat the quantity $\abundnlsp$ as a free
parameter which will be determined experimentally at the LHC.  

\vspace*{\baselineskip}
\noindent {\bf Relic density and the LHC.}\hspace*{0.3cm}
The total abundance of the LSPs is the sum of both thermal and non-thermal 
production contributions
\beqa{
\abundlsp = \abundlsptp + \abundlspntp.
\label{Omega_tot}
}
Since it is natural to expect that the LSP makes up most of CDM in the
Universe, we can re-write the above as
\beqa{
\abundlsptp\left(\treh,\mlsp,\mgluino,\mnlsp,\ldots \right) +
\frac{\mlsp}{\mnlsp} \abundnlsp = \abundlsp = \abundcdm\simeq 0.1.
\label{eq:oh2relation}
}
This is our master formula. 
Once the neutralino NLSP is discovered and its mass is determined at
the LHC with some precision, and so also $\abundnlsp=\abundchi$, then
eq.~(\ref{eq:oh2relation}) will provide a relation between $\treh$ and
$\mlsp$. 

More specifically, in order to evaluate $\abundchi$ at the LHC, the
parameters determining the neutralino mass matrix will have to be
evaluated through measurements at the LHC, as well as the masses of
some other superpartners and Higgs
boson(s)~\cite{oh2atlhc-cmssm,oh2atlhc-mssm}. For our purpose, also
the mass of the gluino $\mgluino$ will have to be known since it
determines the efficiency of E-WIMP production in thermal scatterings.
We assume the usual gaugino mass unification which, at the electroweak
scale implies $\mone=5/3\tsqthw \mtwo = \alpha_1/\alpha_s\mgluino$ (or
$\mone\simeq 0.5\mtwo \simeq  \mgluino/6.5$), where $\mone$ and $\mtwo$
are the bino and wino mass parameters, respectively. (If the gluino
mass is actually measured, we can use it in our expressions
instead.) In this work we assume the lightest neutralino to be mostly a
bino in which case $\mchi\simeq\mone\simeq \mgluino/6.5$.

Note that, if $\abundnlsp$ is determined to be larger than 0.1, the
relic abundance of the LSP may easily agree
with~(\ref{Oh2WMAP}) by taking an appropriate mass ratio
$\mlsp/\mnlsp$ and small enough $\treh$ in order to suppress
TP. Should, on the other hand, $\abundchi$ came out to be smaller than
0.1, a substantial contribution of LSP production in thermal processes
at large enough $\treh$ will be sufficient. Actually, TP may be
dominant even if $\abundchi$ is larger than, or for that matter is
close to, 0.1, depending on the LSP and the neutralino NLSP mass
ratio.

\section{Axino dark matter}
\label{axino}

We first consider the axino as the LSP and the dominant component of
DM in the Universe. As mentioned earlier, for definiteness we take the
NLSP to be the lightest neutralino; the stau case will be considered
below.  It is reasonable to expect that LHC experiments will be able
to probe neutralino (and stau) mass ranges up to some $400\gev$,
depending on other SUSY parameters. Here we are not concerned with
experimental uncertainties of LHC measurements but rather illustrate
the principle of estimating $\treh$ for a given DM candidate in terms
of relation~(\ref{eq:oh2relation}). 

%
\begin{figure*}[t!]
\vspace*{-0.2in} 
\begin{center}
  \begin{tabular}{c c}
    \includegraphics[width=0.45\textwidth]{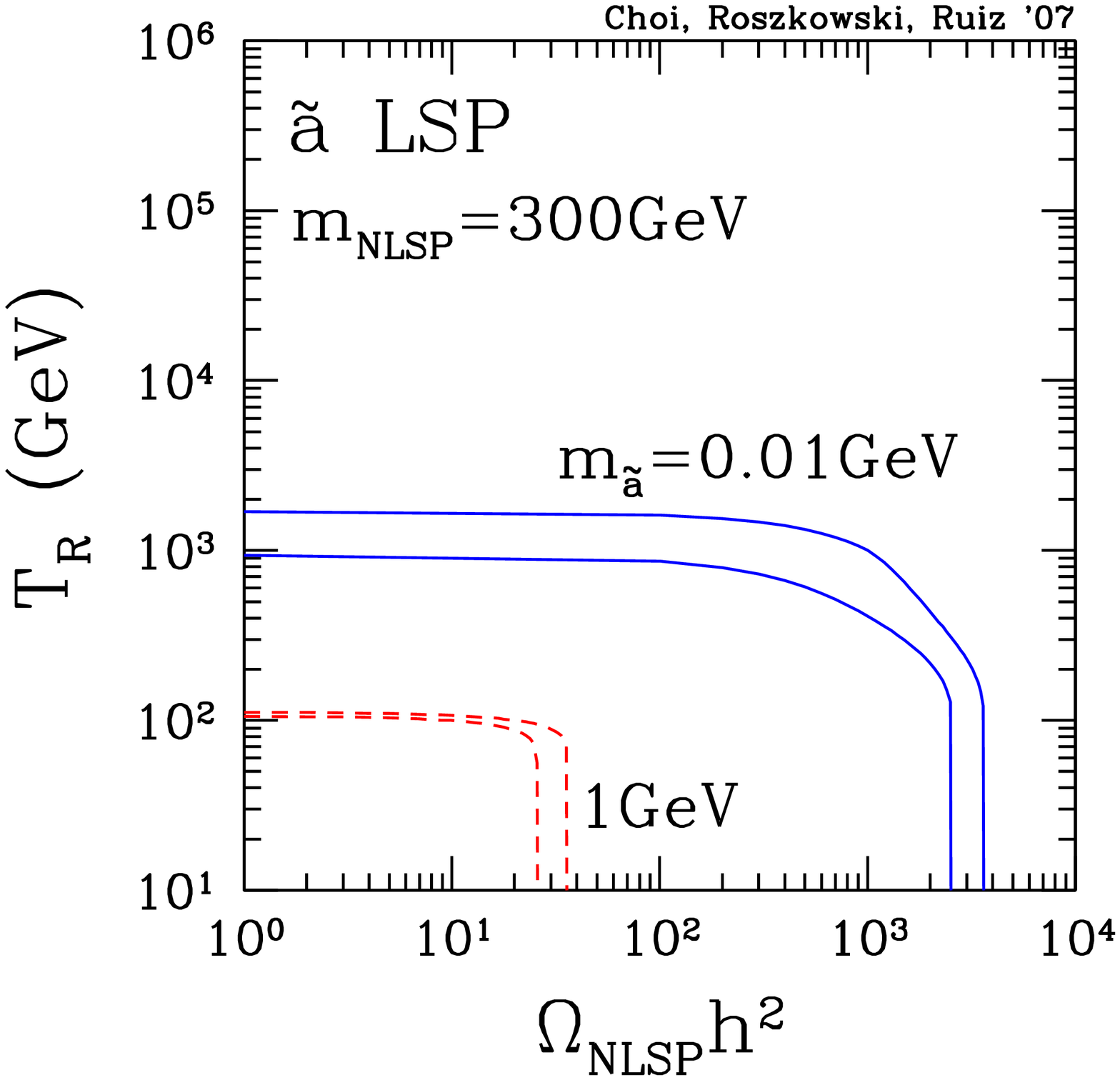}
&
    \includegraphics[width=0.45\textwidth]{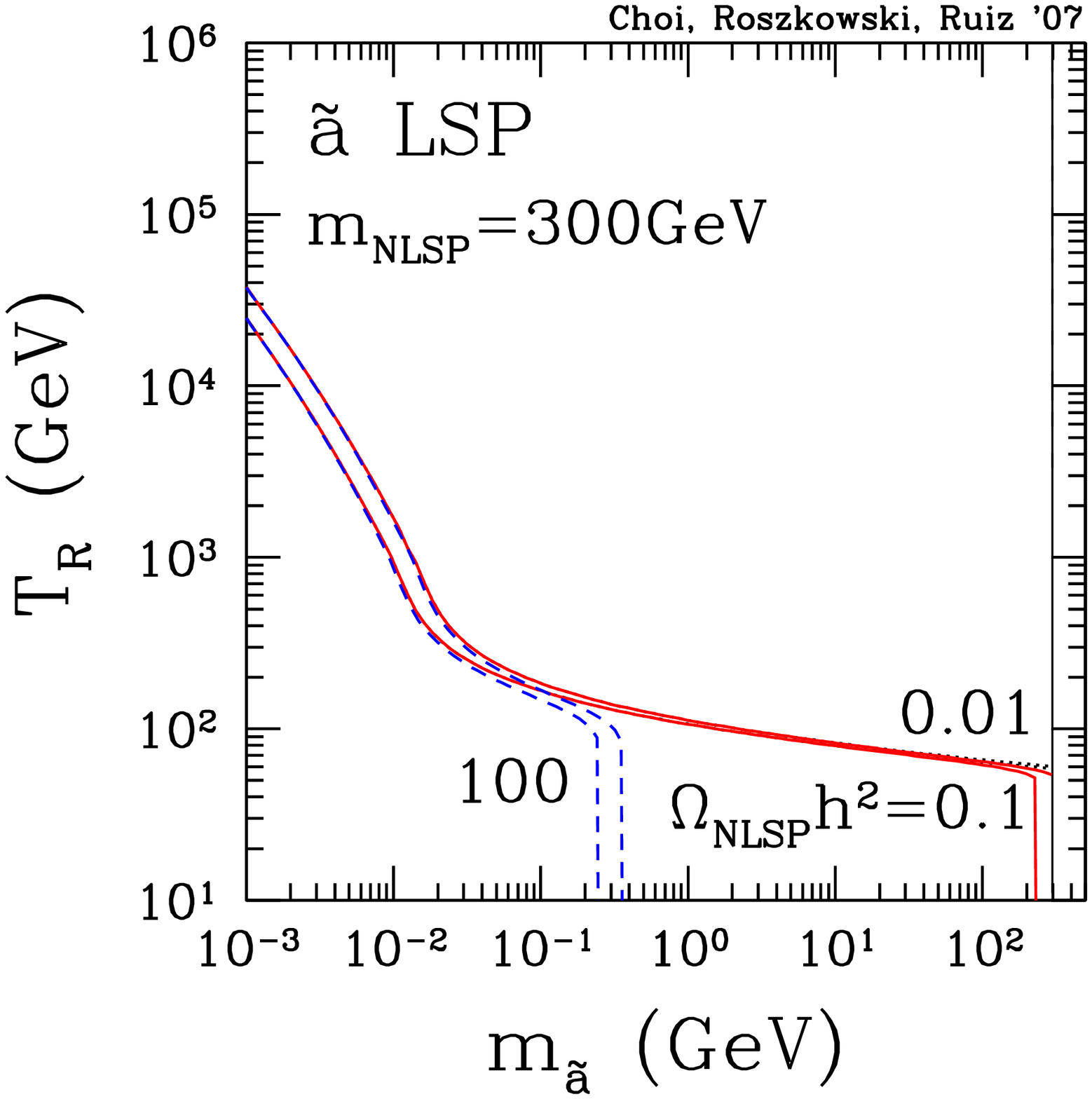}
    \end{tabular}
\end{center}
\caption{Left panel: $\treh$ vs. $\abundnlsp$ for $\mnlsp=300\gev$ and
for $\maxino=0.01\gev$ (solid blue) and $\maxino=1\gev$ (dashed
red). The bands correspond to the upper and lower limits of dark
matter density in~(\protect\ref{Oh2WMAP}).  Right panel: $\treh$
vs. $\maxino$ for $\abundnlsp=100$ (dashed blue), 0.1 (solid red) and
0.01 (dotted black). To the right of the solid vertical line the axino
is no longer the LSP. In both panels we set $\fa=10^{11}\gev$. }
\label{fig:axinotr}
\end{figure*}
\begin{figure}[!tb]
  \begin{center}
    \includegraphics[width=0.45\textwidth]{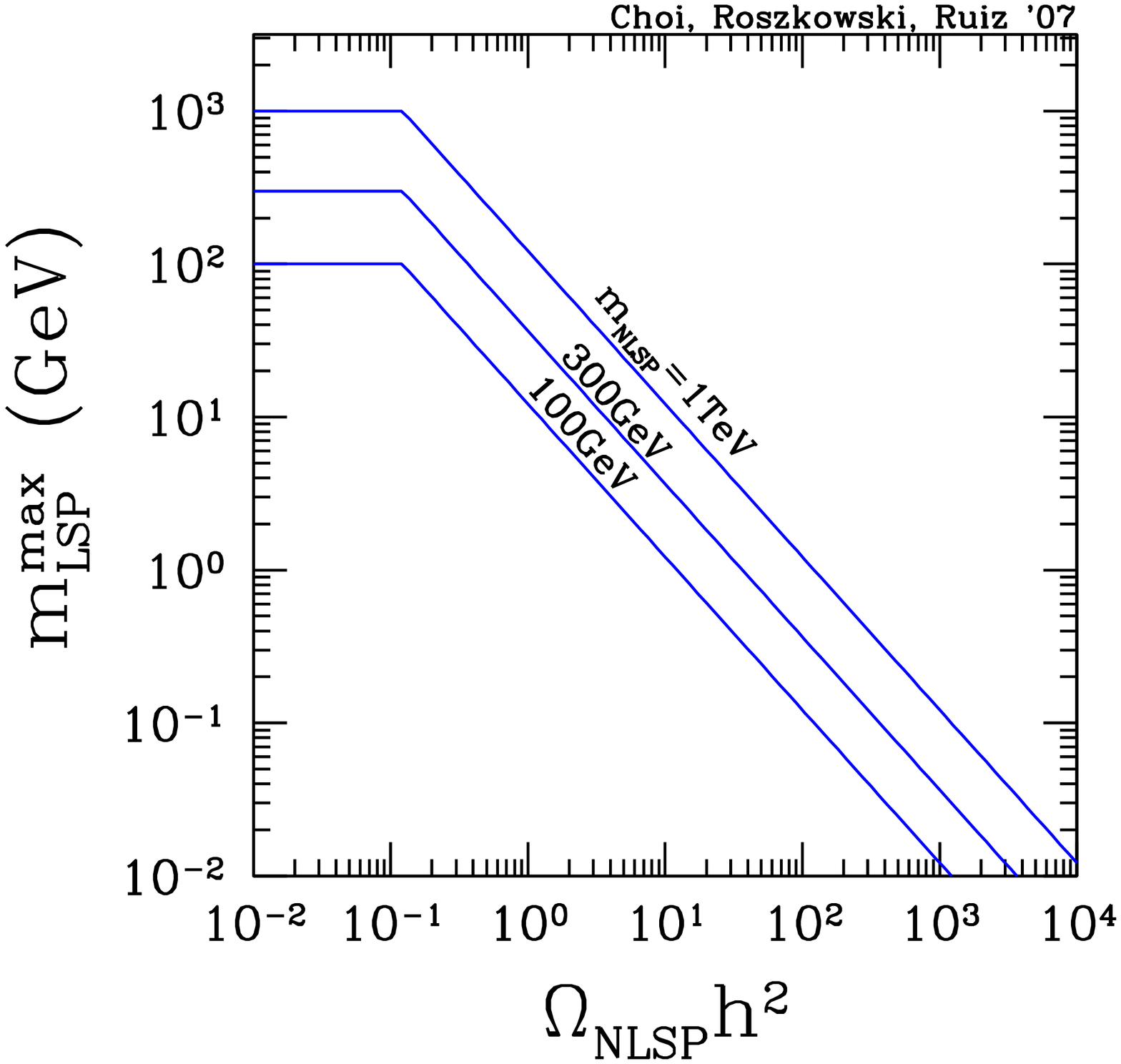}
  \end{center}
\caption{Maximum values of $\mlsp$ as a function of $\abundnlsp$ for
  representative values of $\mnlsp$. Once both $\abundnlsp$ and
  $\mnlsp$ are determined from experiment, the upper bound on $\mlsp$
  can be derived. The plot applies both to the axino and to the
  gravitino LSP.}
\label{fig:mlspmax}
\end{figure}

In the left panel of fig.~\ref{fig:axinotr} we show ranges of
$\abundnlsp$ and $\treh$, such that $\abunda$ is in the
range~(\ref{Oh2WMAP}).
We take a fixed neutralino NLSP
mass $\mnlsp=300 \gev$ and $\maxino=0.01 \gev$ (solid blue) and
$1\gev$ (dashed red). For each choice of the axino mass the two lines
correspond to the upper and lower limits of dark matter density
in~(\ref{Oh2WMAP}).  On the left side non-thermal production is
subdominant while thermal production of axinos gives the main
contribution to their relic density. In this case $\treh$ corresponds
to the value for which $\abundatp \simeq \abundcdm$, and depends on
the axino and NLSP masses since $Y^{\rm
TP}_{\axino}\propto\treh/\fa^2$.  As $\abundantp$ increases,
$\abundatp$ must decrease (in order for the total abundance to remain
close to 0.1) and at some point becomes subdominant. This
point is marked by an abrupt turnover of the contours from
horizontal to vertical.

In the right panel of fig.~\ref{fig:axinotr} we plot $\maxino$
vs. $\treh$ for the same NLSP mass and for $\abundnlsp=100$ (dashed
blue), 0.1(solid red) and 0.01 (dotted black). On the left side,
where $\maxino$ is small, TP dominates, $\abundatp \simeq
\abundcdm$, hence we find
\beqa{
\treh\propto \fa^2/\maxino.
\label{eq:trehmaxinorelation}
}
In this regime $\treh$ is inversely proportional to the axino mass (for a
given $\fa$, see below). (This dependence can be seen analytically by
plugging eq.~(\ref{eq:TP_axino_bs}) into eq.~(\ref{eq:oh2relation})
and taking the limit of negligible contribution from NTP.) The
relation (\ref{eq:trehmaxinorelation}) allows one to derive an {\em
upper bound} on $\treh$ if we use the fact that axinos have to be
heavy enough in order to constitute CDM. Assuming conservatively that
$\maxino\gsim100\kev$~\cite{ckkr} we find
$\trehmax<4.9\times10^5\gev$.

\begin{figure}[!t]
  \begin{center}
  \begin{tabular}{c c}
    \includegraphics[width=0.45\textwidth]{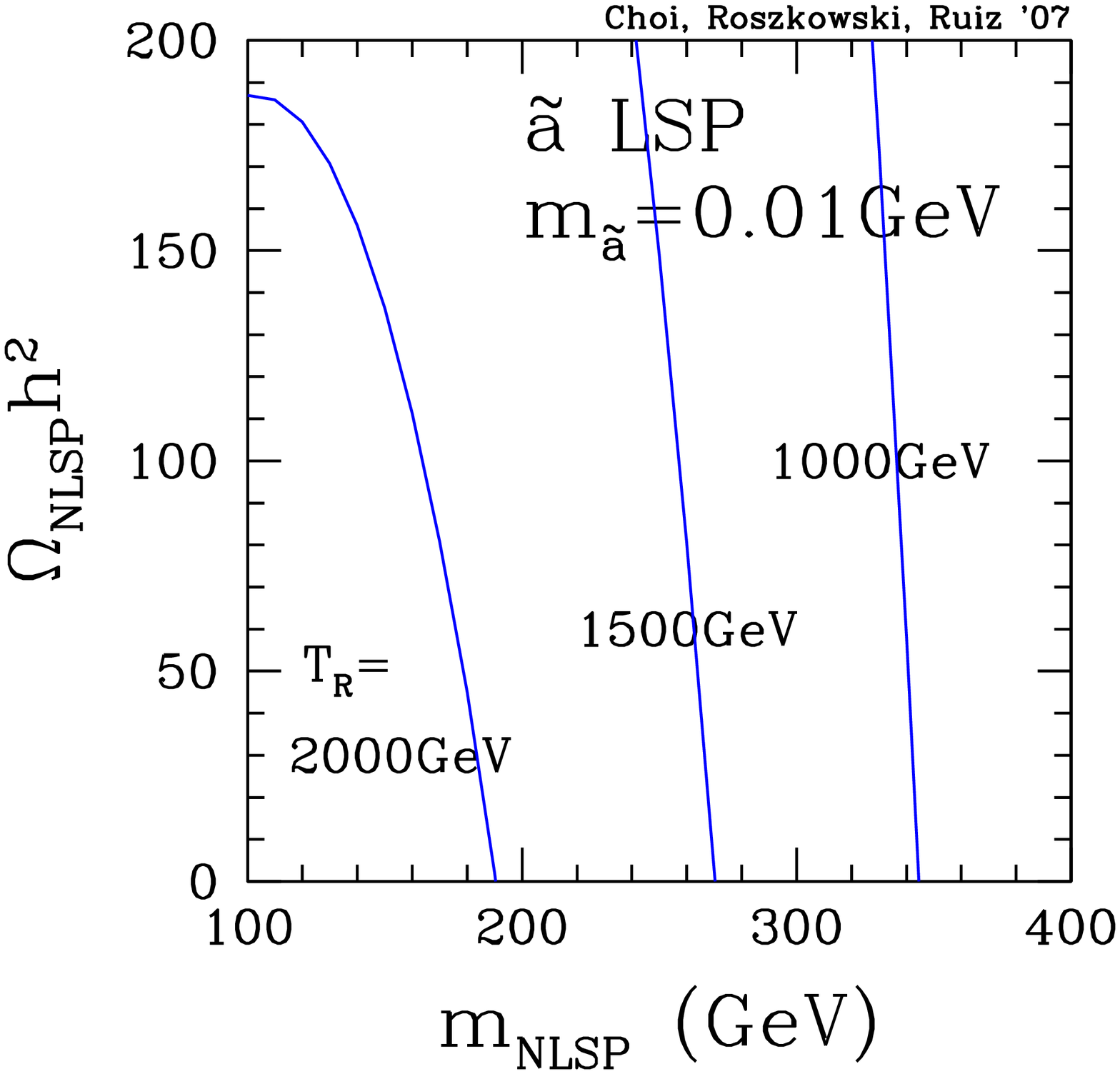}
&
    \includegraphics[width=0.45\textwidth]{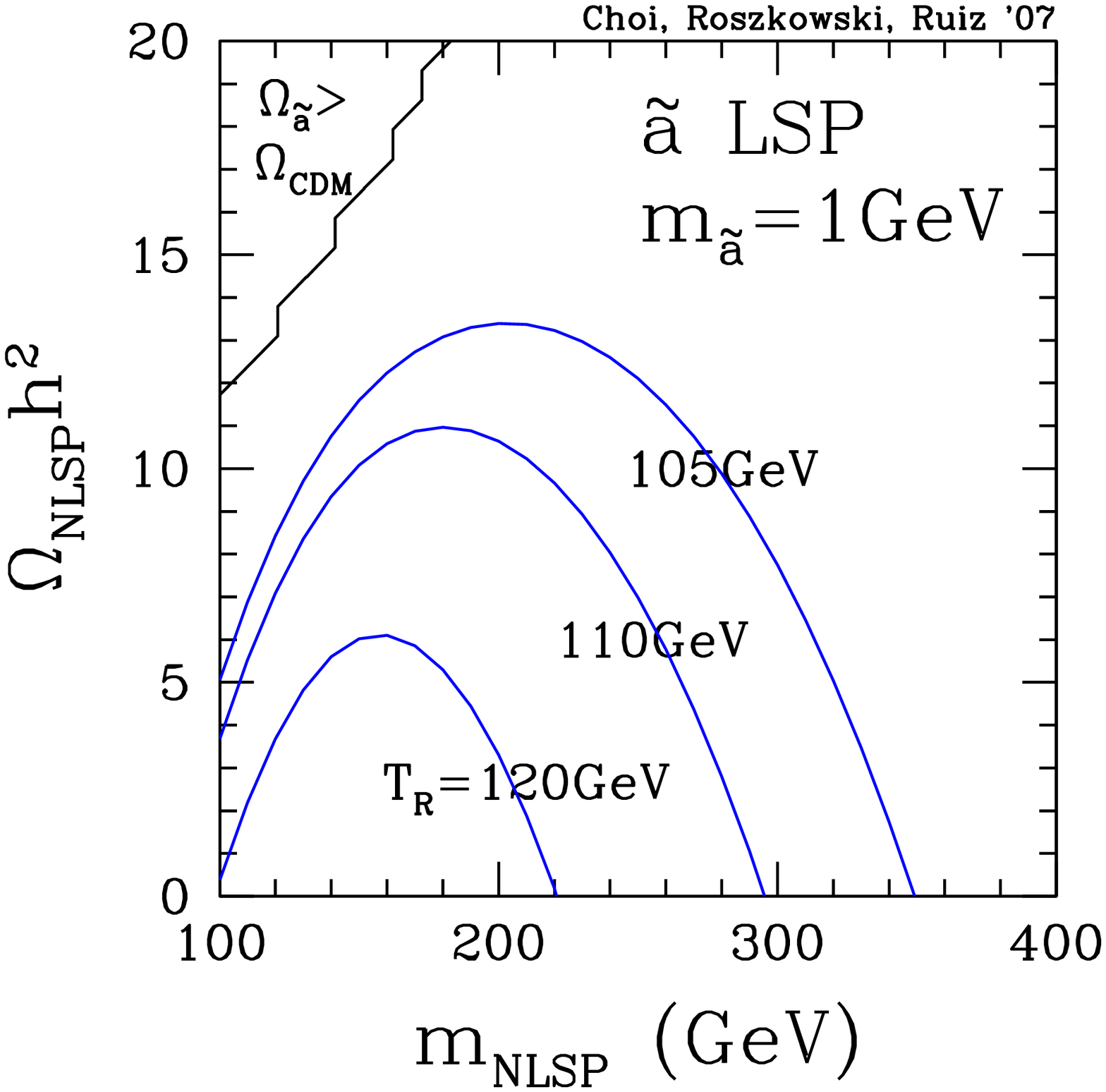}
    \end{tabular}
  \end{center}
\caption{Contours of the reheating temperature in the plane of
  $\mnlsp$ and $\abundnlsp$ such that $\abunda=\abundcdm=0.104$. The axino
  mass is assumed to be  $0.01\gev$ (left panel) and $1\gev$ (right).}
\label{contour_axino1}
\end{figure}

At larger $\maxino$ the NTP contribution becomes dominant and the
dependence on $\treh$ is lost but one can still place a lower bound on
$\treh$.  Since in this regime the LSP mass becomes largest, this
allows one to derive an {\em upper bound} on $\maxino$. This is shown
in fig.~\ref{fig:mlspmax} as a function of $\abundnlsp$ for
$\mnlsp=100,300\gev$ and $1\tev$. Once both $\abundnlsp$ and $\mnlsp$
are determined from experiment, the upper bound on $\mlsp$ can 
be derived. Note that fig.~\ref{fig:mlspmax} actually applies to both
the axino and the gravitino LSP since it follows from
eq.~(\ref{eq:ntp}). On the other hand, the bound is interesting only if
$\abundnlsp\gg0.1$, otherwise it reduces to a trivial condition
$\mlsp<\mnlsp$.

In fig.~\ref{contour_axino1} we plot in the plane of varying $\mnlsp$
and $\abundnlsp$ where contours of $\treh$ for a fixed
$\maxino=0.01\gev$ (left panel) and $1\gev$ (right panel).  In the
left panel, where $\maxino$ is tiny, for sufficiently large $\treh$
thermal production dominates and the dependence on $\abundnlsp$ is
very weak. Indeed, only at small $\maxino$ and large $\abundnlsp$ the
NTP contribution $\propto(\maxino/\mnlsp)\abundnlsp$ starts playing
some role.

\begin{figure*}[t!]
\vspace*{-0.2in} 
  \begin{center}
    \includegraphics[width=0.45\textwidth]{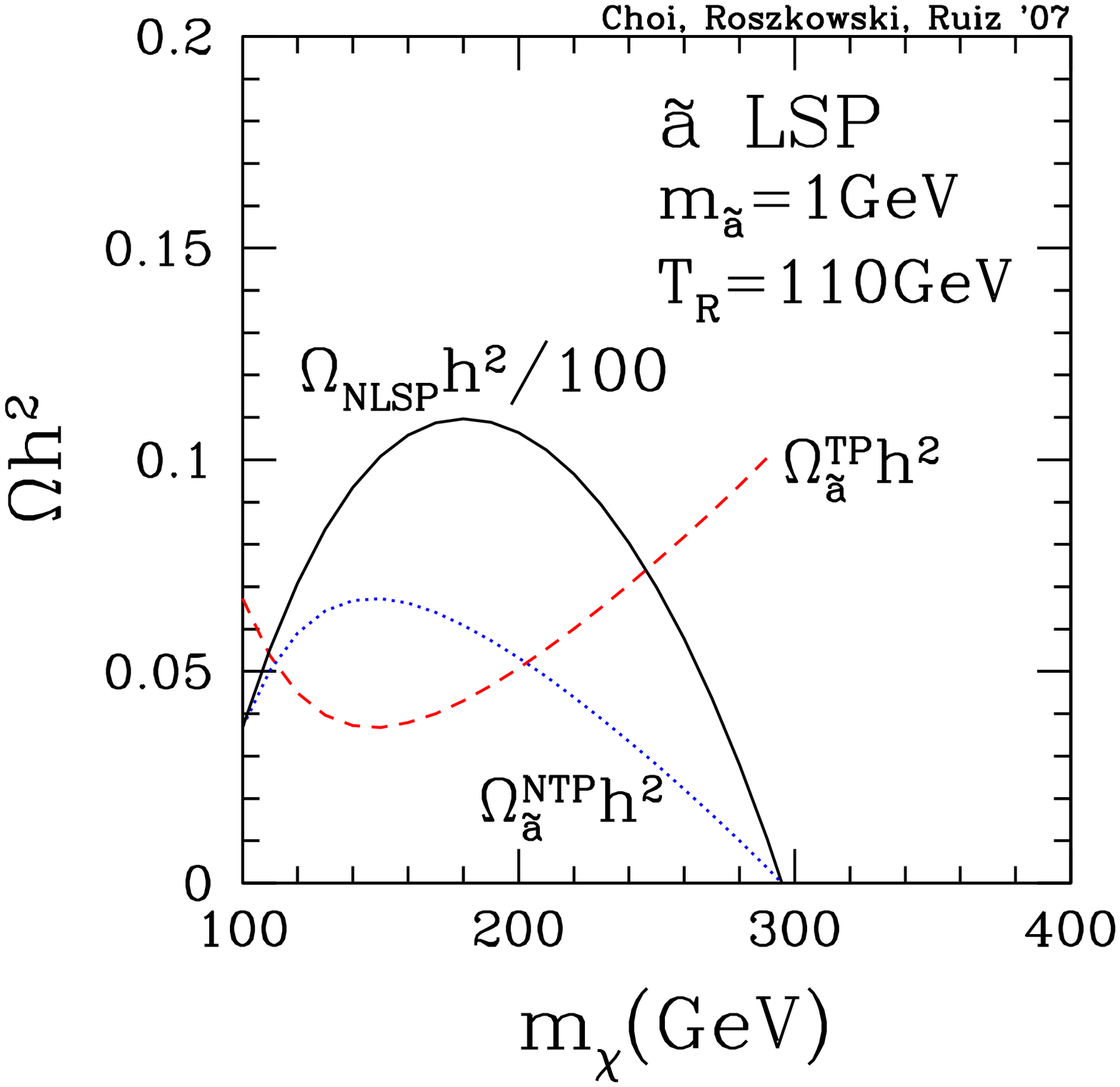}
  \end{center}
\caption{The TP (dashed red) and NTP (dotted blue) contributions to the
  axino relic abundance vs. $\mnlsp=\mchi$ for the case of 
  $\maxino=1 \gev$ and $\treh=110\gev$ from
  fig.~\protect\ref{contour_gravitino1}. The CDM relic abundance has
  been set at its central value of 0.104. Also shown is $\abundnlsp/100$
  (solid black) for this case.}
\label{fig:axinotp+ntp}
\end{figure*}

On the other hand, in the right panel of fig.~\ref{contour_axino1} the
situation is somewhat more complex, with a characteristic turnover
which depends on $\treh$. In order to explain it, in
fig.~\ref{fig:axinotp+ntp} the TP and NTP contribution to the axino
relic abundance are shown explicitly for one of the cases presented
in fig.~\ref{contour_axino1}, along with the corresponding
$\abundnlsp/100$. As the neutralino mass $\mchi=\mnlsp$ increases, the
gluino mass also increases (via the gaugino unification mass
relation). So long as $\mchi\lsim130\gev$, the gluino decay
contribution to $\abundatp$ dominates over the one from squark decays
(with squark masses fixed in here at $1\tev$) and from scatterings
(compare the left panel of fig.~\ref{fig:Y_TP}), but it decreases from
left to right. This is because it is proportional to $\mgluino$ times an
integral over the Boltzmann suppression factor
$\exp{(-\mgluino/T)}$. On the other hand, the squark decay
contribution, which scales as $\left[\mgluino
\left(1-0.05\log(\mgluino/1\tev)\right)\right]^2$,
increases~\cite{crs02}. Overall, as $\mchi$ increases, the TP part of
$\abunda$ first decreases before increasing again. Thus the NTP part
has to first increase and then decrease in order for the sum to remain
constant at 0.104.

In the upper left-hand corner of the right panel of
fig.~\ref{contour_axino1}, for a given ratio of $\maxino/\mnlsp$,
$\abundnlsp$ becomes too large for $\abunda$ not to exceed 0.104,
despite basically turning off the TP contribution (by reducing
$\treh$). As $\treh$ increases, the Boltzmann suppression factor plays
a smaller role and the TP part increases, thus making the NTP part
decrease. This is why, for the same value of $\mnlsp$, the
corresponding value of $\abundnlsp$ decreases with increasing $\treh$,
in agreement with the right panel of fig.~\ref{contour_axino1}.
Generally, for $\maxino=1\gev$ the allowed values of $\treh$ are much
smaller than at smaller $\maxino$ (left panel).

In both figs.~\ref{fig:axinotr} and~\ref{contour_axino1} we can see
that a whole spectrum of values of $\abundnlsp$ coming out from LHC
determinations can be consistent with the axino relic abundance
satisfying $\abunda\simeq0.1$. They can range from much smaller to
those much larger than the ``WMAP range''~(\ref{Oh2WMAP}), depending
on the mass of the LSP. Moreover, $\abundnlsp$ does not have to take
largest values in the NTP-dominated case, as one would initially
expect. On the other hand, when $\mnlsp$ is much larger than $\mlsp$,
NTP typically dominates, especially if $\abundnlsp$ is also large. On
the other hand, if $\mnlsp$ is not much larger than $\mlsp$ and
$\abundnlsp$ is also on the small side, then TP typically
dominates. While this dependence is not rigorous, it may help one distinguishing
between the TP and NTP-dominated regimes once $\mnlsp$ and
$\abundnlsp$ are experimentally determined.

Finally we comment on the dependence on $\fa$. Its value is limited to
lie above some $10^8\gev$ from cooling of red giants and below some
$10^{12}\gev$ from inflation if it took place prior to the
Peccei-Quinn transition. Otherwise larger values are
allowed~\cite{wilczek07-golden}.  In our numerical examples we have
set $\fa=10^{11}\gev$. Since only TP of axinos depends on $\fa$ (as
$\abundatp\propto 1/\fa^2$) while NTP does not, in
figs.~\ref{fig:axinotr} and~\ref{contour_axino1} the regions
dominated by TP will be affected accordingly. In other words,
$\treh\propto \fa^2/\maxino$, as stated above.

\section{Gravitino dark matter}
\label{gravitino}

The case of gravitino LSP, with its relic abundance
$\abundg$ satisfying the range~(\ref{Oh2WMAP}), can be analyzed in an
completely analogous way, except for two important aspects. 

The first has to do with Big Bang Nucleosynthesis (BBN). After the
NLSPs freeze-out from thermal plasma and start decaying into the LSPs,
energetic photons and hadronic jets are also produced. If this takes
place during or after BBN, such particles may destroy successful
predictions of standard BBN of light element abundances.
In the axino LSP case, the NLSP decay occurs normally before BBN epoch
begins~\cite{ckr,ckkr}.  Thus the axino LSP is mostly
free from BBN constraints.  However, with gravitino couplings to
ordinary matter being so much weaker, NLSP decays to gravitinos take place
much later, varying from $1\sec$ to $10^{12}\sec$, depending on the
masses of the two.  Thus BBN usually provides severe constraints
on the yield of the NLSP.

The neutralino NLSP is almost excluded with $\mgravitino\gsim 1 \gev$
due to BBN~\cite{fengetal,ccjrr}.  On the other hand, if the gravitino
mass is less than $\sim1 \gev$, the lifetime of the neutralino becomes
smaller than about $1 \sec$ and the window of neutralino NLSP opens up
again~\cite{ccjrr}.  In our numerical examples below we will not
impose the constraint from BBN but will indicate in which case it is
satisfied or not.

\begin{figure*}[t!]
\vspace*{-0.2in} 
  \begin{center}
  \begin{tabular}{c c}
    \includegraphics[width=0.45\textwidth]{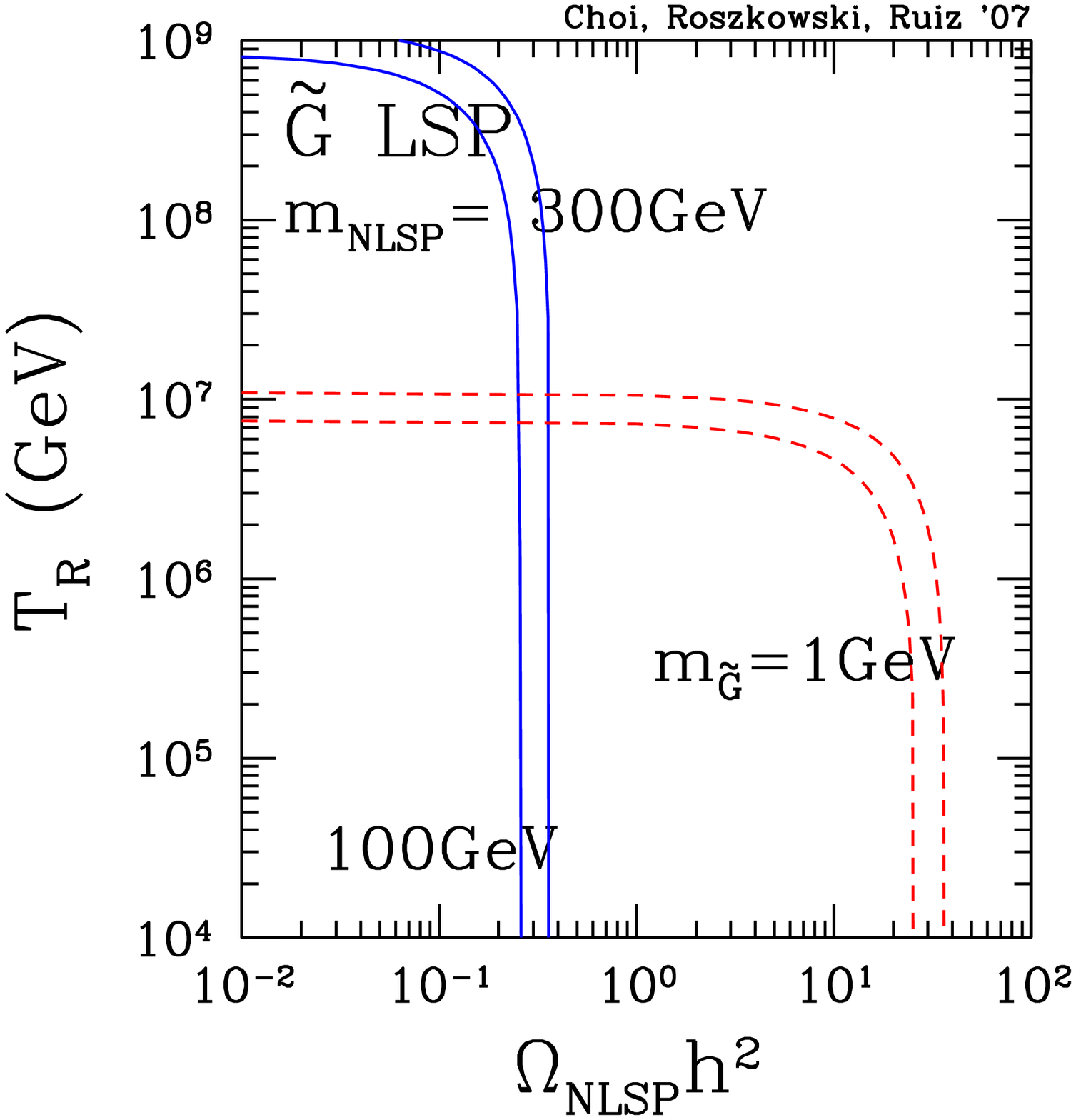}
&
    \includegraphics[width=0.45\textwidth]{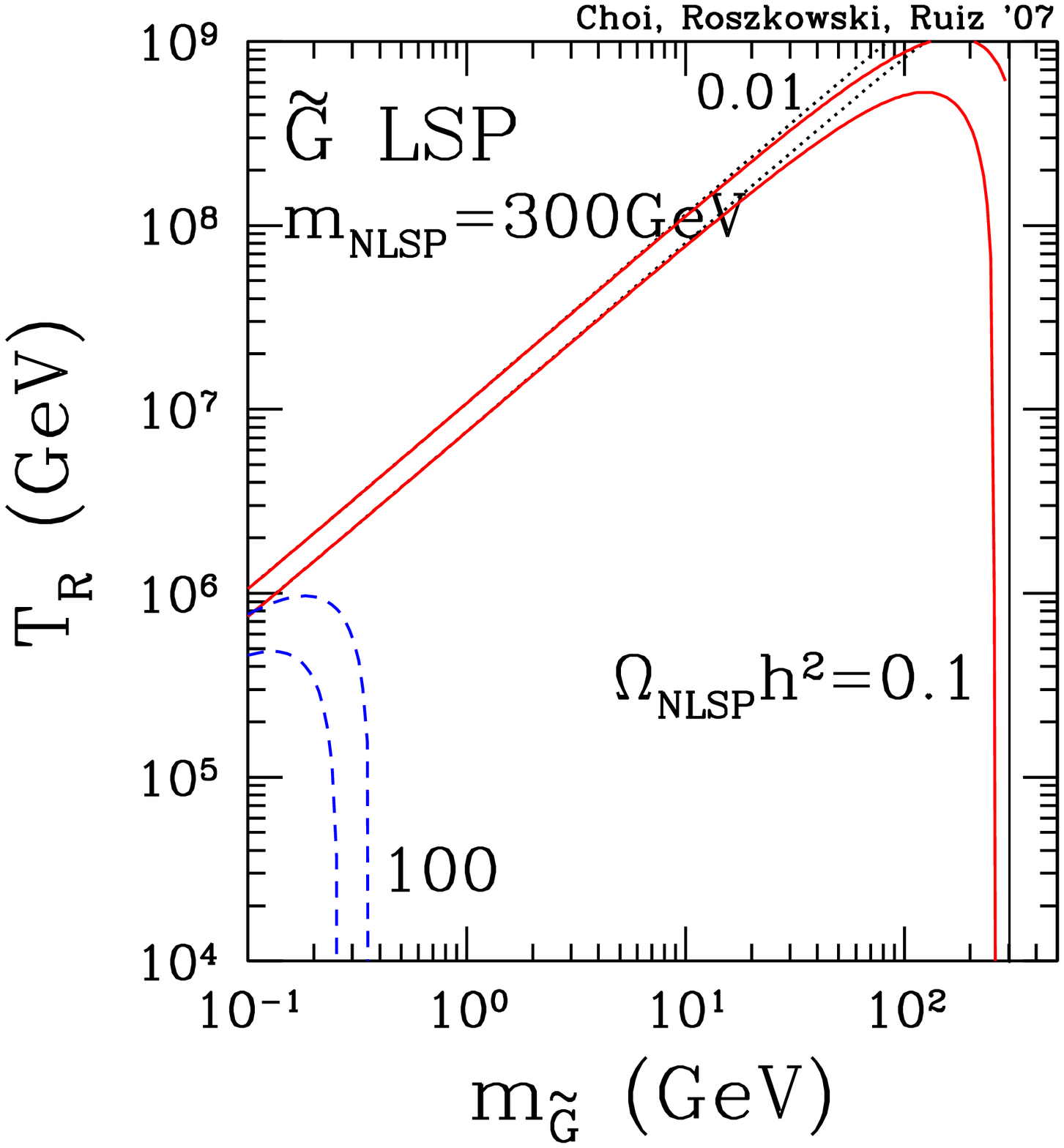}
    \end{tabular}
  \end{center}
\caption{Left panel: $\treh$ vs. $\abundnlsp$ for $\mnlsp=300\gev$ and
for $\mgravitino=0.01\gev$ (solid blue) and $\mgravitino=1\gev$
(dashed red). The bands correspond to the upper and lower limits of
dark matter density in~(\protect\ref{Oh2WMAP}).  Right panel: $\treh$
vs. $\mgravitino$ for $\abundnlsp=100$ (dashed blue), 0.1 (solid red)
and 0.01 (dotted black).  To the right of the solid vertical line the
gravitino is no longer the LSP.}
\label{fig:gravitinotr}
\end{figure*}

The second important difference with the axino case was already
discussed in sec.~\ref{production}. While $\yaxinotp$ is independent
of the axino mass, in the  gravitino case $\ygravitinotp\propto
1/\mgravitino^2$. Thus $\abundatp\propto \maxino\treh$ while
$\abundgtp\propto \treh/\mgravitino$. In other words, if TP dominates,
$\abundgtp\simeq0.1$, we find
\beqa{
\treh\propto\mgravitino.
\label{eq:trehmgravitinorelation}
}
This is reflected in the right panel of fig.~\ref{fig:gravitinotr}
where we show $\treh$ vs. $\mgravitino$ for our nominal value of the
NLSP mass of $300\gev$ and for some fixed values of $\abundnlsp$. So
long as $\mgravitino\ll\mnlsp$ and $\abundnlsp$ is small enough, e.g.
$\abundnlsp=0.1$ (denoted by a solid red curve), or less, the gravitino relic
density is dominated by TP and $\treh$ scales linearly with
$\mgravitino$ so that $\abundatp\simeq 0.1$. On the other hand, when
$\abundnlsp=100$ (dashed blue), NTP becomes rapidly dominant even with
sub-GeV LSP mass.  This can also be seen in the left panel of
fig.~\ref{fig:gravitinotr} where we show $\treh$ vs. $\abundnlsp$ for
the same NLSP mass and for $\mgravitino=0.01\gev$ (solid blue) and
$\mgravitino=1\gev$ (dashed red). 
For the smaller value of $\mgravitino$ the
TP contribution remains dominant up to much larger allowed values of
$\abundnlsp$ (horizontal part of the curves) before finally NTP takes
over (vertical part).  Note that, just like with the axino LSP case
before, once we know $\mnlsp$ and $\abundnlsp$, we can place an {\em
upper bound} on the mass of the gravitino assumed to be the dominant
component of CDM in the Universe, see fig.~\ref{fig:mlspmax}.

%
\begin{figure}[!t]
  \begin{center}
    \includegraphics[width=0.45\textwidth]{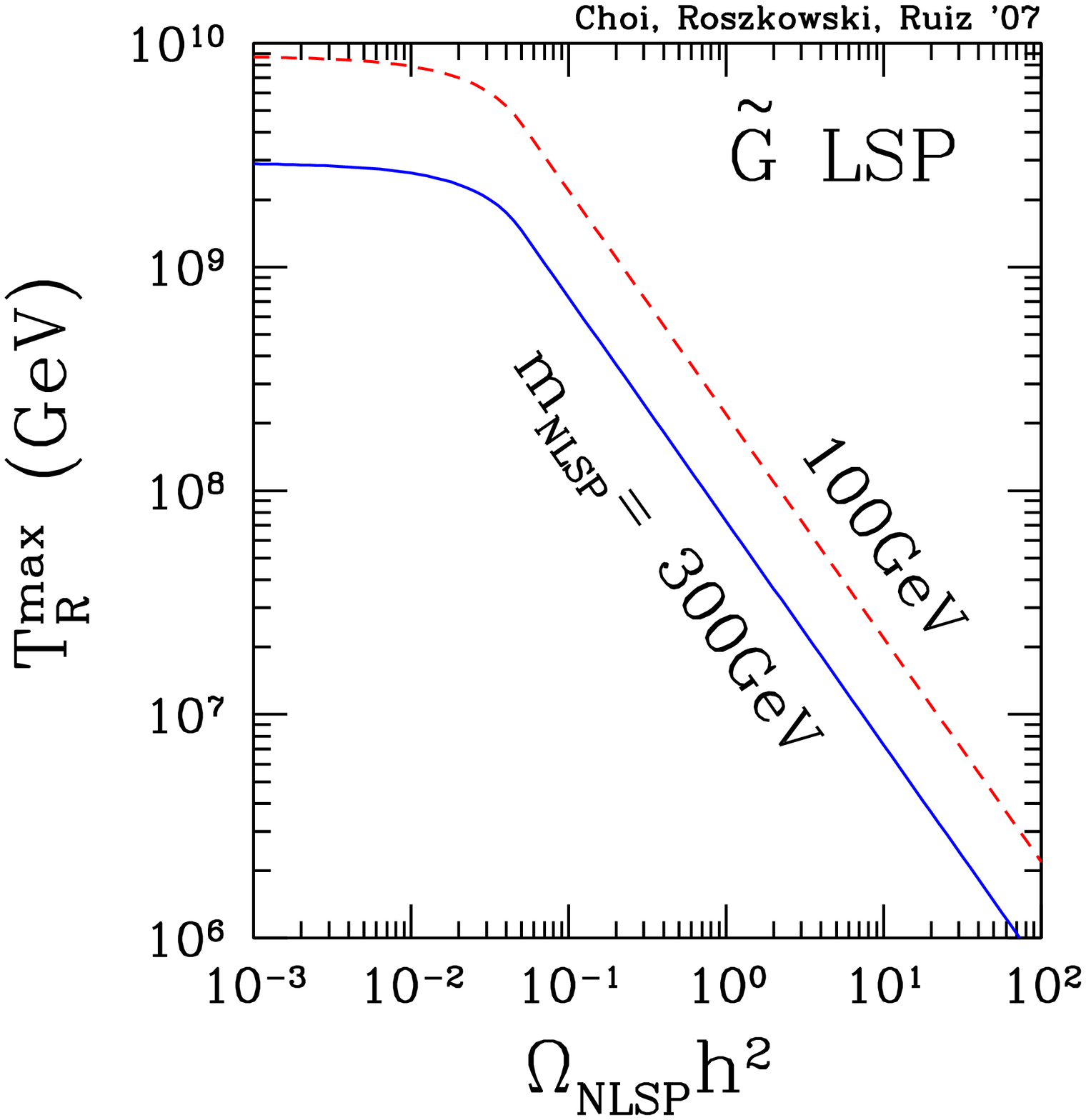}
  \end{center}
\caption{Maximum reheating temperature $\trehmax$ vs. NLSP relic density 
$\abundnlsp$ with gravitino DM for NLSP mass $\mnlsp = 100 \gev$
  (dashed red) and $300 \gev$ (solid blue).}
\label{fig:trehmaxvsonlsp}
\end{figure}
%

The turnover between the TP and NTP dominance allows one to derive a conservative {\em
upper bound} $\trehmax$ which, unlike for the axino CDM, even without
knowing the gravitino mass. This can be seen from
eqs.~(\ref{eq:abundgbbb})--(\ref{Omega_tot}) from which an expression
for $\treh$ as a quadratic function of $\mgravitino$ can easily be
found. Its maximum can then be expressed as
\beq \trehmax=\left\{
\begin{array}{ll}
\frac{10^{8}\gev}{0.27}\left(\frac{1\tev}{\mgluino}\right)^2\left(\frac{\mnlsp}{100\gev}
\right)\frac{1}{4\abundnlsp}, & \quad\textrm{for}\ \abundnlsp > 0.05\\ 
\frac{10^{10}\gev}{0.27}\left(\frac{1\tev}{\mgluino}\right)^2\left(\frac{\mnlsp}{100\gev}
\right)\left( 0.1-\abundnlsp \right), & \quad\textrm{for}\ \abundnlsp < 0.05 
\end{array}\right., 
\label{eq:trehmaxgrav}
\eeq
where we have used $\abundlsp=0.1$. This is plotted in
fig.~\ref{fig:trehmaxvsonlsp} for two representative choices of
$\mnlsp=100,300\gev$, the former one  being roughly the lowest value
allowed by LEP.

\begin{figure*}[t!]
\vspace*{-0.2in} 
\begin{center}
  \begin{tabular}{c c}
    \includegraphics[width=0.45\textwidth]{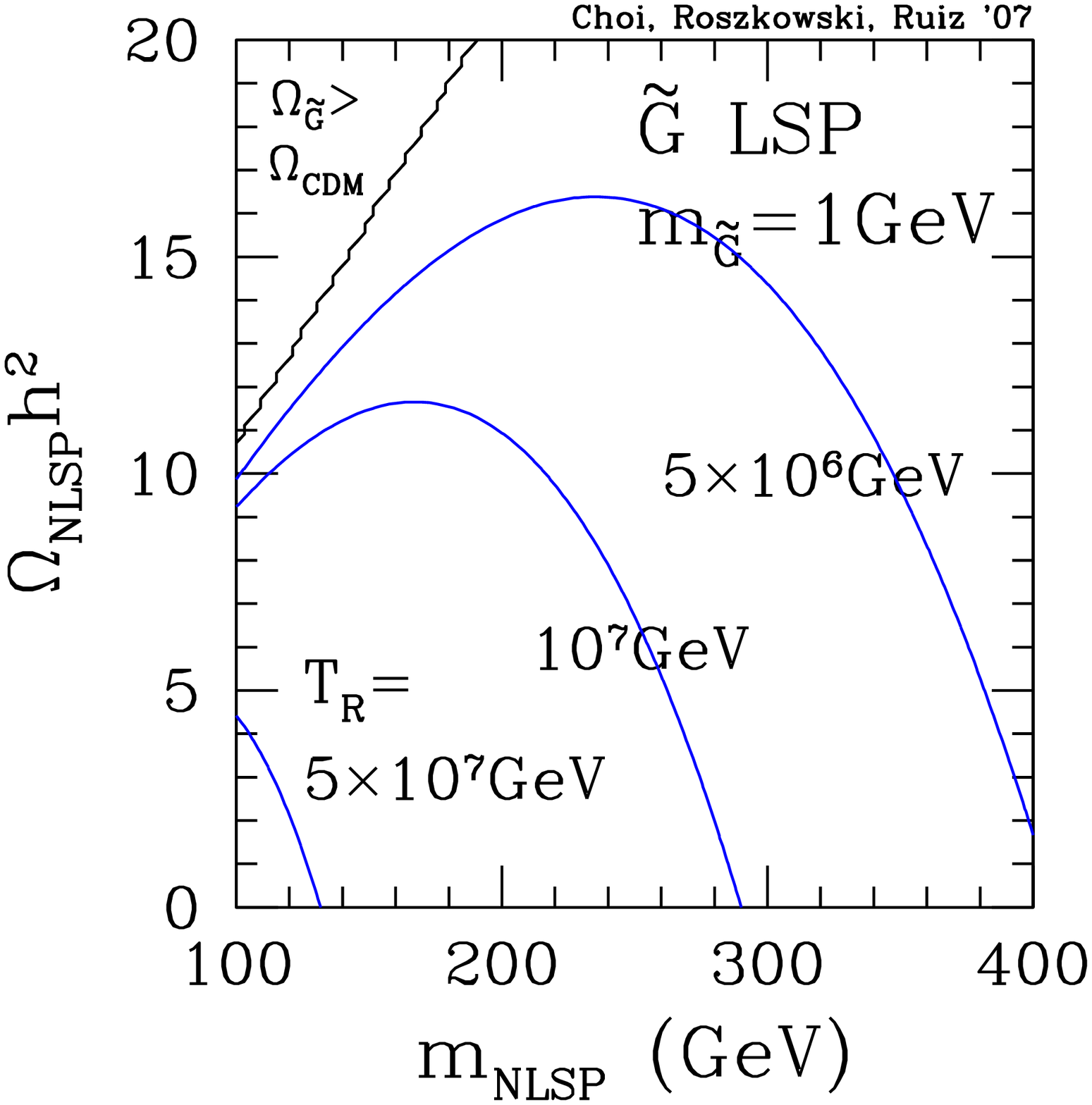}
&
    \includegraphics[width=0.45\textwidth]{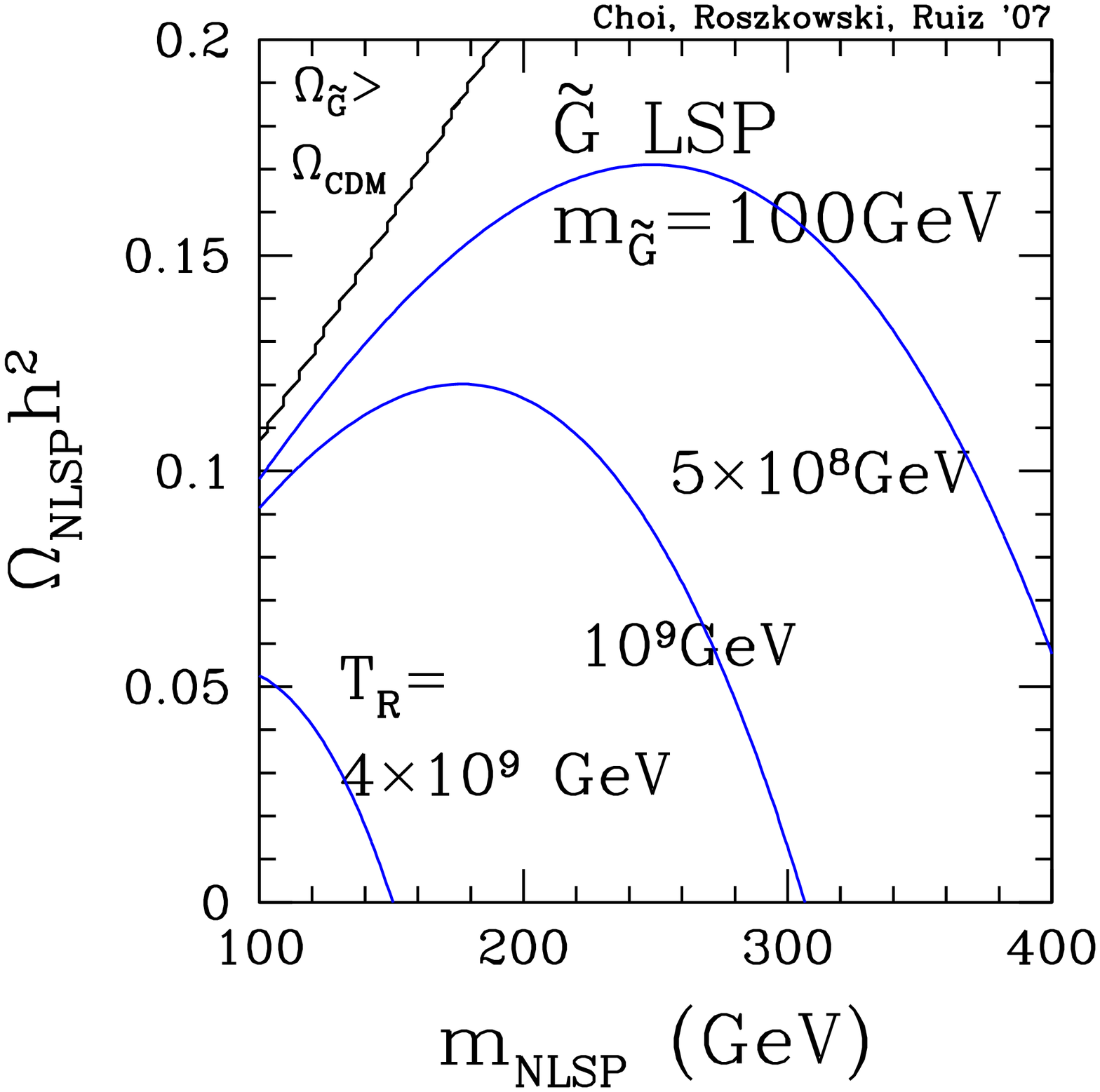}
    \end{tabular}
\end{center}
\caption{Contours of the reheating temperature in the plane of
  $\mnlsp$ and $\abundnlsp$ such that $\abundg=\abundcdm=0.104$. The gravitino
  mass is assumed to be  $1\gev$ (left panel) and $100\gev$ (right panel).}
\label{contour_gravitino1}
\end{figure*}

\begin{figure*}[t!]
\vspace*{-0.2in} 
  \begin{center}
    \includegraphics[width=0.45\textwidth]{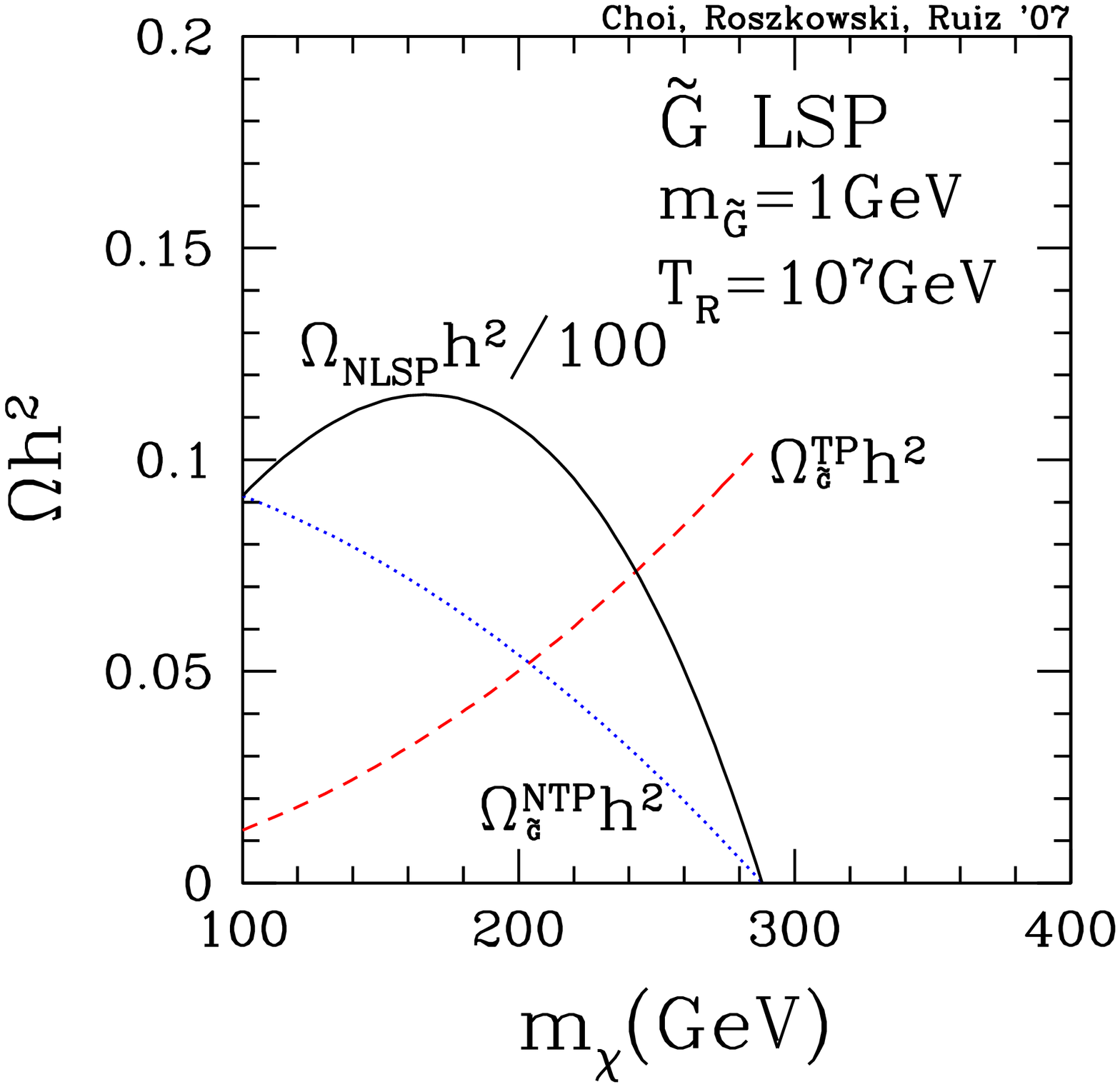}
  \end{center}
\caption{The TP (dashed red) and NTP (dotted blue) contributions to
  the gravitino relic abundance vs. $\mnlsp=\mchi$ for the case of
  $\mgravitino=1\gev$ and $\treh=10^7 \gev$ from
  fig.~\protect\ref{contour_gravitino1}. The CDM relic abundance has
  been set at its central value of 0.104. Also shown is $\abundnlsp/100$
  (solid black) for this case.}
\label{fig:gravitinotp+ntp}
\end{figure*}

In fig.~\ref{contour_gravitino1} we present, in analogy with
fig.~\ref{contour_axino1}, contours of $\treh$ for a fixed
$\mgravitino=1\gev$ (left panel) and $100\gev$ (right panel). In both,
at sufficiently large $\abundnlsp$ we find a similar turnover of the
contours as in the right panel of fig.~\ref{contour_axino1} for the
axino case.  Again, in order to elucidate the situation, in
fig.~\ref{fig:gravitinotp+ntp} the TP and NTP contributions to the
gravitino relic abundance are explicitly shown for one of the cases
presented in fig.~\ref{contour_gravitino1}, along with the
corresponding $\abundnlsp/100$, in analogy with
fig.~\ref{fig:axinotp+ntp} for the axino case. The (initially
subdominant) TP part increases with increasing $\mgluino$ (which grows
with $\mchi$), thus the NTP part has to decrease in order for the sum
to remain constant at 0.104. In the upper left-hand corner of both
panels of fig.~\ref{contour_gravitino1}, for a given ratio of
$\mgravitino/\mnlsp$, $\abundnlsp$ becomes too large and $\abundg$
exceeds 0.104, despite basically turning off the TP contribution (by
reducing $\treh$).

In the gravitino LSP case it is easy to derive a simple expression for
the value of the NLSP mass at which the turnover in $\abundnlsp$ in
fig.~\ref{contour_gravitino1} takes place. By taking the NLSP to be
the bino (as we do in our numerical examples), plugging the expression
for $\abundgtp$ from eq.~(\ref{eq:abundgbbb}) into
eq.~(\ref{eq:oh2relation}) and making a simplifying assumption
$\mgluino(\mu)\simeq \mgluino\simeq 6.5\mchi$ one obtains an
expression for $\abundchi$ as a cubic function of $\mchi$. By putting
its first derivative to zero one arrives at
\begin{equation}
\mchi\simeq 54\, \left(\frac{10^{10}\gev}{\treh}\right)^{1/2}
\left(\frac{\mgravitino}{10^2\gev}\right)^{1/2}\gev, 
\label{eq:gturnover}
\end{equation}
which adequately reproduces the positions of the peaks in
fig.~\ref{contour_gravitino1}.

The case of
$\mgravitino=1\gev$ is on the borderline of being consistent with BBN,
while the mass of $100\gev$ is not (if the NLSP is the neutralino). On
the other hand, both cases can be allowed for some choices of MSSM
parameters if the NLSP is the stau, see below.
Note that in the gravitino case resulting values of $\treh$
are typically significantly larger than for axinos. On the other hand, for
sufficiently small $\mgravitino\lsim1\gev$ (roughly consistent with BBN
constraints for NLSP neutralino) we find $\treh\sim10^{6-7}\gev$,
while for MeV-mass axinos $\treh\sim10^{4-5}\gev$ (compare the right
panel of fig.~\ref{fig:axinotr}), which are of similar order.

\section{Stau NLSP}
\label{staunlsp}

Finally, we discuss the case of the lighter stau as the NLSP. Its
couplings are of a sufficiently similar size to those of the
neutralino that, for the same mass, its freeze-out temperature and
yield are basically the same, hence  is also its NTP contribution to either
axino or gravitino relic abundance. For this reason, our
numerical results presented in
figs.~\ref{fig:axinotr}--\ref{contour_gravitino1} for the neutralino
NLSP case apply directly to the stau NLSP case.\footnote{One caveat is
that the gluino mass, which is important in TP, is not directly
related anymore to the stau mass.} 

What is different is the impact of BBN constraints. Like with the
neutralino, stau decays result in both electromagnetic and hadronic
showers. Axino production from NLSP decays mostly takes place before
the period of BBN, thus this case remains basically unconstrained. On
the hand, the gravitino LSP case is again strongly constrained.
Detailed analyses have shown that BBN constraints are somewhat weaker
with stau NLSP than for the neutralinos~\cite{fengetal,rrc04,ccjrr}. As a
result, the LSP gravitino mass in the range of tens, or even hundreds,
of~GeV, can remain allowed and consistent with $\abundg$ in the ``WMAP
range''.  Stau NLSP is also disallowed if its lifetime is longer than
about $10^3\sec$, as otherwise it would form metastable bound states
with nuclei which would affect BBN abundances of light elements, as
recently pointed out in~\cite{Pospelov:2006sc}, see
also~\cite{Bird:2007ge,Kaplinghat:2006qr}.  The stau lifetime is
dominated by two-body decay to tau and gravitino which, neglecting tau
mass, is given by~\cite{fengetal}
\beqa{
\tau (\stau \rightarrow \tau +\gravitino)
=6.1\times10^3 \sec \left(\frac{1\tev}{\mstau}\right)^5 
\left(\frac{\mgravitino}{100 \gev}\right)^2
\left(1-\frac{\mgravitino^2}{\mstau^2}\right)^{-4}.
}
In our numerical example in the right panel of
fig.~\ref{fig:gravitinotr}, with $\mstau=300\gev$, where we have also
taken $\mchi=477\gev$, the condition $\tau_{\stau}>10^3\sec$ implies
$\mgravitino\lsim2\gev$ and $\treh\lsim 9\times 10^6\gev$. Increasing
$\mstau$ to $1\tev$ and $\mchi$ to $1.5\tev$ leads to
$\mgravitino\lsim40\gev$ and $\treh\lsim 4\times 10^8\gev$. 

The main point is that, with the neutralino NLSP, BBN constraints
basically exclude gravitino CDM with the mass larger than about
$1\gev$, while in the stau NLSP case much larger values can be
allowed.

\section{Summary}
\label{summary}

If the axino or the gravitino E-WIMP is the lightest superpartner and
the dominant component of CDM in the Universe, then, under favorable
conditions, it will be possible to determine the reheating temperature
after inflation from LHC measurements. Basically, one will need to
measure the neutralino mass and other parameters determining its
``relic abundance'' - a non-trivial and highly exciting possibility in
itself. The quantity may be found to be different from the ``WMAP
range''~(\ref{Oh2WMAP}), thus suggesting a non-standard CDM candidate.
Additionally, one will need to know the mass of
the gluino, either through a direct measurement, or via the gaugino
unification mass relation, since it plays an important role in E-WIMP
production at high $\treh$.  The neutralino is likely to be
sufficiently long-lived to appear stable in LHC detectors, but in the
early Universe it would have decayed into the axino or the gravitino LSP. Since each
of them is extremely weakly interacting, it would be hopeless to look
for them in usual WIMP dark matter searches.

In the axino LSP case, if thermal production dominates, a consistency of
its relic abundance with the ``WMAP range''~(\ref{Oh2WMAP}) implies
$\treh\propto \fa^2/\maxino$. If non-thermal production is dominant
instead, then, for a known value of $\mchi$, the value of $\abundchi$
will imply an upper bound on $\maxino/\fa^2$, which in turn will imply
a lower bound on $\treh$.

For the gravitino LSP, in the thermal production dominated case the analogous
dependence is $\treh\propto \mgravitino$. If non-thermal production
dominates then the value of $\abundchi$
will imply an upper bound on $\mgravitino$. 
For both axino and gravitino one can express this as
\beq
\mlspmax=\min\{\mnlsp, \frac{\abundcdm}{\abundnlsp}\mnlsp\},
\label{eq:mlspmax}
\eeq
which is independent of $\treh$.

For both relics, one can also derive an upper
bound on $\treh$, although in the axino case has to further assume
that it is cold DM. 
On the other hand, for both axino and gravitino, it will be difficult to
distinguish between the TP and NTP regimes but some information may be
helpful. Typically, when $\mchi$ is large and $\abundnlsp$ is also large
then NTP is dominant, while if $\abundnlsp$ is small then TP typically
dominates.

Similar conclusions as for the neutralino (for both E-WIMPs) apply if the NLSP is 
the stau, instead. Additionally, the very discovery of a charged,
massive and (seemingly) stable state at the LHC will immediately imply
that dark matter in the Universe is not made up of the usual neutralino. 

With so many similarities between both E-WIMPs, a natural question
arises whether it would be possible to distinguish them in a collider
experiment. Unfortunately, with neutralino NLSP this is unlikely to be
possible. On the other hand, with the stau NLSP being electrically charged,
enough of them could possibly be accumulated. By comparing their
3-body decays, it may be possible to only tell whether Nature has
selected the axino or the gravitino as the lightest superpartner, but
to actually also determine their mass with some reasonable
precision~\cite{Buchmuller:2004rq,bchrr05}. Needless to say, this
would allow one to determine also the reheating temperature.

In this work, we focused on demonstrating the principle of determining
$\treh$ at the LHC, and did not concern ourselves with uncertainties
of experimental measurements. These are likely to be substantial at
the LHC, the case we have mostly focused on here, unless one considers
specific models~\cite{bk05}. A more detailed study will be required to
assess their impact. An analogous study can be performed in the
context of the planned Linear Collider. We also implicitly assumed
that only one relic dominates the CDM component of the Universe, which
in principle does not have to be the case. For example, the axion and
the axino could co-exist and contribute comparable fractions to the
CDM relic density. (Note that this could be the case also in the
``standard'' case of the stable neutralino.) This would introduce an
additional uncertainty and would lead to replacing the derived values
of $\treh$ with an upper bound on the quantity.

\medskip
{\bf Acknowledgements} \\ This work is supported by an STFC
grant. K.-Y.C. has been partially supported by a PPARC grant, by the
Ministerio de Educacion y Ciencia of Spain under Proyecto Nacional
FPA2006-05423 and by the Comunidad de Madrid under Proyecto HEPHACOS,
Ayudas de I+D S-0505/ESP-0346. L.R is partially supported by the EC
6th Framework Programmes MRTN-CT-2004-503369 and
MRTN-CT-2006-035505. R.RdA is supported by the program ``Juan de la
Cierva'' of the Ministerio de Educaci\'{o}n y Ciencia of Spain.  The
author(s) would like to thank CERN for hospitality and support and the
European Network of Theoretical Astroparticle Physics ENTApP ILIAS/N6
under contract number RII3-CT-2004-506222 for support. This project
benefited from the CERN-ENTApP joint visitor's programme on dark
matter, 5-9 March 2007 and the CERN Theory Institute ``LHC-Cosmology
Interplay'', 25 June - 10 Aug 2007.

\newcommand\jcap[3] 
		{{\it J.\ Cosmol.\ Astrop.\ Phys.\ }{\bf #1} (#2) #3}
\newcommand\mnras[3]   { 
		{{\it Mon.\ Not.\ R.\ Astron.\ Soc.\ }{\bf #1} (#2) #3}}
\newcommand\apjs[3]   { 
		{{\it Astrophys.\ J.\ Supp.\ }{\bf #1} (#2) #3}}
\newcommand\aipcp[3]   { 
		{{\it AIP Conf.\ Proc.\ }{\bf #1} (#2) #3}}


\end{document}